\documentclass[english,tightenlines,eqsecnum,amsmath,amssymb,aps,prd,superscriptaddress,nofootinbib,showpacs,floats]{article}
\usepackage{amsmath,amsfonts,amssymb,multicol}
\usepackage{graphicx,caption,subcaption}
\usepackage{epstopdf}
\usepackage{cite}
\usepackage[usenames]{xcolor}
\usepackage{epstopdf}
\definecolor{AHZ}{rgb}{0.0,1,0.0}
\usepackage[colorlinks,linkcolor=AHZ,citecolor=red,bookmarks]{hyperref}
\usepackage{rotating}
\usepackage[mathscr]{eucal}
\usepackage{graphicx}
\usepackage[T1]{fontenc}
\usepackage{babel}
\usepackage{authblk}
\usepackage{epsfig}
\usepackage{color}
\usepackage{amssymb}
\usepackage{fancyhdr}

\def\nn{\nonumber\\}
\newcommand{\f}[2]{\frac{#1}{#2}}
\def\be{\begin{equation}}
\def\ee{\end{equation}}
\def\bea{\begin{eqnarray}}
\def\eea{\end{eqnarray}}

\def\bwt{\begin{widetext}}
	\def\ewt{\end{widetext}}

\usepackage{amsmath}

\usepackage{amssymb}

\begin{document}
	
\title{Casimir Wormholes in Brans-Dicke Theory}
\author{${^1}$Amir Hadi Ziaie\thanks{ah.ziaie@maragheh.ac.ir}}
\author{${^2}$Mohammad Reza Mehdizadeh\thanks{mehdizadeh.mr@uk.ac.ir}}
\affil{{\rm ${^1}$Research~Institute~for~Astronomy~and~Astrophysics~of~ Maragha~(RIAAM), University of Maragheh,  P.~O.~Box~55136-553,~Maragheh, Iran}}
\affil{{\rm ${^2}$Department~of~ Physics,~ Shahid~ Bahonar~ University, P.~ O.~ Box~ 76175, Kerman, Iran}}
\renewcommand\Authands{ and }
\maketitle
\begin{abstract}
In recent years there has been a growing interest in the field of wormhole physics in the presence of Casimir effect. As this effect provides negative energy density, it can be utilized as an ideal candidate for the exotic matter required for creating a traversable wormhole. In the context of modified theories of gravity such as Brans-Dicke (BD) theory~\cite{BDTH}, wormhole geometries {have} been vastly investigated{. However}, the scientific literature is silent on the issue of BD wormholes in the presence of Casimir energy. Our aim in the present study is to seek for static spherically symmetric solutions representing wormhole configurations in BD theory with Casimir energy as the supporting matter. The Casimir {setup} we assume comprises two electrically neutral, infinitely large parallel planes placed in a vacuum. We then consider the Casimir vacuum energy density of a scalar field in such a configuration with Dirichlet {and} mixed boundary conditions. In the former case the corresponding Casimir force is attractive and in the latter this force is repulsive. We present exact zero tidal force wormhole solutions as well as those with non vanishing redshift function for both types of Casimir energies. The conditions on wormhole solutions along with the weak (WEC) and null (NEC) energy conditions put constraints on the values of BD coupling parameter. These constraints are also subject to the value of BD scalar field at the throat and the throat radius. We therefore find that BD wormholes in the presence of Casimir energy can exist without violating NEC and WEC (for the repulsive Casimir force). Finally, we examine the equilibrium condition for stability of the obtained solutions using Tolman-Oppenheimer-Volkoff (TOV) equation.
\end{abstract}
\maketitle
\section{Introduction}
Undoubtedly, one of the most fascinating features of general relativity ({\rm GR}) that has attracted many researchers so far, is the possible existence of hypothetical geometries which have nontrivial topological structure, known as wormholes. Such structures constitute a short-cut or tunnel between two universes or two separate regions of the same universe~\cite{FLoboBook,khu}. Wormhole physics can originally be traced back to Ludwig Flamm in 1916 who tried to give interpretations of the Schwarzschild solution through 2D embedding diagram of {\it Schwarzschild wormhole}~\cite{Flamm}. In 1935 specific wormhole-type solutions were considered by Einstein and Rosen~\cite{ERose}. Their motivation was to construct potential models for elementary particles represented by a \lq{}\lq{}{\it bridge}\rq{}\rq{} {which connect} two identical sheets. Such a mathematical description of a physical space that is connected by a wormhole-type solution was subsequently denoted as the \lq{}\lq{}{\it Einstein-Rosen bridge}\rq{}\rq{}. After the pioneering work by Einstein and Rosen, studies in the field of wormhole physics were laid dormant for about two decades. It was at the end of 1950s when Misner and Wheeler through their pioneering works~\cite{misner-wheeler,Wheelerworm} invented the concept of wormhole in order to provide a mechanism for having charge without charge. They realized that wormholes connecting two asymptotically flat spacetimes could provide possible existence of non-trivial solutions to the coupled Einstein-Maxwell field equations, where, the lines of electric field flux as observed in one part of the universe could thread the throat and reappear in other part. These objects were denoted as \lq{}\lq{}geons\rq{}\rq{}, i.e., gravitational-electromagnetic entities~\cite{misnerwheelerworks}. However, it was later realized that the geonlike-wormhole structures seem to be a mere curiosity since they were understood as non-{traversable}~\cite{FulWheel} wormholes and additionally would develop some type of singularity~\cite{GerochJMath}, see also~\cite{hisworm} for historical notes. 
\par
Over the past decades, there has been a remarkable amount of literature produced {in the realm} of wormhole physics. This renewed interest in the subject was mainly sparked through the seminal paper by Morris and Thorne~\cite{mt,mt1}. In the framework of GR, they introduced static spherically symmetric metric and discussed the required conditions for physically meaningful Lorentzian traversable wormholes. The Morris-Thorne (MT) wormholes allow a two way communication between two regions of the spacetime by a minimal surface called the wormhole throat, through which, matter and radiation can travel freely. However, traversability of MT wormhole requires inevitably the violation of {\rm NEC}. In other words, in order that the throat of these wormholes can be kept open some form of \lq{}\lq{}{\it exotic matter}\rq{}\rq{} is needed~\cite{mt1}. This type of matter has negative energy density and its energy momentum tensor ({\rm EMT}) violates the {\rm NEC}~\cite{khu,khu1}. Although the violation of energy conditions, at least in classical GR, is unacceptable from the common viewpoint of physicists, several researches have tried to address the issue of exotic matter in wormhole physics. Work along this line has been performed in the literature for different matter distributions, for example, traversable wormhole geometries constructed by matter fields with exotic {\rm EMT} have been studied in~\cite{MarcoChianese2017}. The case of phantom or quintom-type energy as the ingredient to sustain traversable wormholes has been investigated in~\cite{phantworm} and wormholes supported by nonminimal interaction between dark matter and dark energy has been explored in~\cite{intdarksec}, see also~\cite{lobocgqreview} for a comprehensive review. At quantum level, it has been shown that some effects due to quantum field theory, e.g., Hawking evaporation process~\cite{Hawevp} and Casimir effect~\cite{Caseffect} can allow for violation of energy conditions. Negative energy densities in wormhole configurations can also be produced through gravitational squeezing of the vacuum~\cite{negendensqueez}, see also~\cite{khu,Klinkhammer1991} for more details. However, due to the problematic nature of exotic matter, many attempts have been made so far in order to minimize its usage with the aim of overcoming the issue of energy conditions in wormhole structures~\cite{minexot}. Also, in order to avoid the usage of exotic matter in traversable MT wormholes, several modifications to GR have been proposed~\cite{modsgrex}. More precisely, the need for exotic matter in classical MT wormholes can be resolved in modified theories of gravity which for example contain higher curvature corrections or fields other than metric in their gravitational actions. These corrections or additional degrees of freedom that are absent in GR can support traversable wormhole spacetimes. In this regard, several studies have been performed in recent years among which we can quote: traversable wormholes in higher-dimensional GR~\cite{higherdimw}, nonsymmetric gravitational theory~\cite{nonsymgr}, Einstein-Gauss-Bonnet gravity~\cite{gmfl}, brane world scenarios~\cite{braneworm}, Lovelock~\cite{LOVEWORM} and $f({\rm R})$ gravity theories~\cite{fr}, Einstein-Cartan theory~\cite{ecworm}, modified gravities with curvature-matter coupling~\cite{Garcia-Lobo}, Rastall gravity~\cite{rastallworm} and other theories~\cite{otherworms}.
\par
Among different matter sources supporting wormhole geometries, Casimir effect is a well-known source that owing to its associated negative energy, it was considered as a good candidate to produce a traversable wormhole, a long time ago by Morris and Thorne~\cite{mt1} and sometime later by Visser~\cite{khu}. However, only very recently a traversable wormhole solution with Casimir energy as its supporting matter has been found by Garattini~\cite{Garaworm}. After this innovative work, many researchers have investigated wormhole configurations endowed with Casimir energy in different frameworks, such as, corrections due to generalized uncertainty principle~\cite{GUPCAS}, modified gravity theories~\cite{CASWMODG}, Casimir source modified by a Yukawa term~\cite{YukCas}{, Casimir wormholes in Yang-Mills theory~\cite{YMCASW}} and other works~\cite{Casotherworks}.
\par
The search for exact spacetimes admitting wormhole structures in BD theory has opened up a new field of study in the literature. In particular, the occurrence of classical Lorentzian wormholes in BD theory is quite intriguing since it is a natural theory that emerged as a Machian alternative to GR. The motivation for studying these structures in BD theory is that this theory involves a scalar field that can play the role of classical exotic matter required for construction of traversable Lorentzian wormholes. The search for wormhole spacetimes in BD theory has been initiated by Agnese and La Camera~\cite{bd} where they obtained a static spherically symmetric vacuum solution that supports a two-way traversable wormhole for $\omega<-2$ and one way for $\omega>-3/2$. Nandi et al.~\cite{bd1} also showed that three of the I-IV classes of BD solutions obtained by Brans~\cite{brans1962} support a two-way traversable wormhole for $\omega<-2$ and $0<\omega<\infty$. Analytical wormhole solutions in BD theory in the presence of matter were found by Anchordoqui et al.~\cite{bd2}. The authors obtained some regions of the parameter space in which the BD scalar field may play the role of exotic matter, implying that it might be possible to build a
wormhole-like spacetime in the presence of ordinary matter at the throat. Paging through the literature, one finds many interesting
articles on BD wormholes today, see for instance~\cite{bd3,bd3rev,bd4,bd5,bd6,bd7}. Nevertheless, in the scientific literature, there is no trace of the consequences of Casimir energy on wormholes in BD theory. Motivated by these considerations, in the present work we investigate wormhole solutions in this theory in the presence of Casimir energy. Our report is then is organized as follows: In Sec.~(\ref{BDFES}) we give a brief review on BD theory along with its corresponding field equations. In Sec.~(\ref{wormsolssec}), using the general static spherically symmetric MT metric we recast the BD field equations, given in tensorial form, into a set of coupled differential equations. In subsection (\ref{redzero}) we search for exact wormhole solutions with zero tidal force and check the energy conditions for them. Subsection (\ref{nonzerored}) is devoted to the solutions with non-vanishing redshift function. At the end of this subsection we examine stability of the obtained solutions through equilibrium condition. Our conclusions are drawn in Sec.~(\ref{concluding}). We use a system of units so that $\hbar=c=G=1$.
\par
\section{Brans-Dicke theory: Action and Field Equations}\label{BDFES}
The BD action in Jordan frame is given by~\cite{Faraoni}
\bea S=\f{1}{16\pi}\int  d^4x\sqrt{-g}\left[\phi R-\f{\omega}{\phi}g^{\mu\nu}\nabla_\mu\phi\nabla_\nu\phi-V(\phi)\right]+S^{(m)},\label{actionbd}
\eea
where
\bea 
S^{(m)}=\int  d^4x\sqrt{-g}\mathcal{L}^{(m)},
\eea
is the action describing ordinary matter fields, i.e., any form of matter different from scalar field $\phi$, and $\omega$ is a dimensionless coupling parameter. We note that matter is not directly coupled to the BD scalar field as $\mathcal{L}^m$ is independent of $\phi$. However, in the gravitational sector we have a non-minimal coupling between {the} scalar field and curvature. In BD theory, the gravitational interaction is described by the metric tensor and by the BD scalar field, which together with the matter variables describe the dynamics of the model~\cite{fujimaeda}. Varying action (\ref{actionbd}) with respect to metric we obtain the BD field equation as
\bea\label{bdfield}
G_{\mu\nu}=\f{8\pi}{\phi}T^{(m)}_{\mu\nu}+\f{\omega}{\phi^2}\Big[\nabla_\mu\phi\nabla_\nu\phi-\f{1}{2}g_{\mu\nu}\nabla^\alpha\phi\nabla_\alpha\phi\Big]+\f{1}{\phi}\left(\nabla_\mu\nabla_\nu\phi-g_{\mu\nu}\Box\phi\right)-\f{V(\phi)}{2\phi}g_{\mu\nu},
\eea
where
\bea\label{emt}
T^{(m)}_{\mu\nu}=-\f{2}{\sqrt{-g}}\f{\delta}{\delta g^{\mu\nu}}\left(\sqrt{-g}\mathcal{L}^{(m)}\right),
\eea
is the matter EMT. Variation of the action with respect to $\phi$ gives
\be\label{bdevolve}
\f{2\omega}{\phi}\Box\phi+R-\f{\omega}{\phi^2}\nabla^\alpha\phi\nabla_\alpha\phi-\f{dV}{d\phi}=0.
\ee
Now, taking the trace of Eq.~(\ref{bdfield}) we get the following expression for Ricci scalar 
\be\label{bdevolve1}
R=-8\pi\f{T^{(m)}}{\phi}+\f{\omega}{\phi^2}\nabla^\alpha\phi\nabla_\alpha\phi+\f{3\Box\phi}{\phi}+\f{2V}{\phi},
\ee
where $T^{(m)}$ is the trace of EMT. Using the above equation to eliminate $R$ from Eq.~(\ref{bdevolve}) we finally get the evolution equation for BD scalar field as
\be\label{bdevol}
(2\omega+3)\Box\phi-8\pi T^{(m)}-\phi\f{dV}{d\phi}+2V=0.
\ee
In view of the above equation we observe that the BD scalar field is not directly coupled to $T^{(m)}_{\mu\nu}$ or $\mathcal{L}^m$, it acts back on ordinary matter only through the metric tensor in the manner determined by Eq.~(\ref{bdfield}). The BD coupling parameter is a free parameter of the theory which represents the strength of coupling between the scalar field and the spacetime metric. In the present work, we exclude the case with $\omega=-3/2$ as for this value of coupling parameter, the BD theory is known to be pathological~\cite{BDpath} in the sense that the BD scalar field is non-dynamical and, more importantly the Cauchy problem becomes ill-posed~\cite{omega32}. 
\section{Wormhole Structures}\label{wormsolssec}
In the present section we seek for wormhole configurations in BD theory whose supporting energy density originates from {the} Casimir effect. The associated energy to this effect in various configurations has been calculated for flat and curved boundaries~\cite{miltonbook}. Studies conducted in this field over the past years have shown that the Casimir energy and its corresponding attractive or repulsive force can assume different forms under various circumstances, namely, $i)$ geometry and distance of the objects involved in a Casimir {setup}, $ii)$ the type of the quantum field under consideration e.g., electromagnetic, fermionic and scalar field, and $iii)$ the boundary conditions imposed on the objects, {e.g., Dirichlet and Neumann boundary conditions for a scalar field~\cite{miltonbook,pcp}, conductor boundary conditions for the electromagnetic field~\cite{advances} and bag boundary conditions for the spinor field~\cite{spinorcas}}. For example, in the case of scalar field confined within a cubic configuration the Casimir energy density is given as $\rho_{\rm c}=-0.015/a^4$ where $a$ is the cuboid side length~\cite{cuboidcas}. For the case of wedge-shaped geometries, i.e., the situation where space is divided by two conducting planes making an angle $\varphi_0$ with each other, the associated energy density for electromagnetic field~\cite{emwedge} and scalar field~\cite{sfwedge} is given respectively as
\bea\label{wedgecas}
\rho_{\rm c}(r)=-\f{1}{720\pi^2r^4}\left(\f{\pi^2}{\varphi_0^2}-1\right)\left(\f{\pi^2}{\varphi_0^2}+11\right),~~~~~~~~~~~~\rho_{\rm c}(r)=-\f{1}{1440\pi^2r^4}\left(\f{\pi^4}{\varphi_0^4}-1\right),
\eea
where $r$ is the radial distance from the intersection of two planes. For a sphere of radius $R$ the Casimir energy density of the electromagnetic field has been calculated and the result is $\rho_{\rm c}(r)\approx {\rm const}/RD^3$ where $D=R-r\ll R$ and $r$ is the distance from the center of the sphere~\cite{cassphere,cassphere1}. Work along this line has been carried out for various other geometries with more complicated configurations such as, scalar Casimir energy for two spheres and a sphere and a plane~\cite{scassphp}, the case of a cylindrical shell~\cite{cynshellcas}, two parallel cylinders and a cylinder parallel to a plane~\cite{cascynp}, see also~\cite{miltonbook,advances,milton2004} for more details. In view of the above considerations, one may recognize that the Casimir energy is proportional to inverse powers of distance between the objects involved in a Casimir configuration. We therefore take the energy density associated to Casimir effect as $\rho_{\rm c}(r)=\lambda/r^m$ where $\lambda$ and $m>0$ are constant parameters. 
\par
Let us now consider the general static and spherically symmetric line element representing a wormhole spacetime which is given by 
\begin{align}\label{metric}
	ds^2=-{\rm e}^{2\Phi(r)}dt^2+\left(1-\f{b(r)}{r}\right)^{-1}dr^2+r^2d\Omega^2,
\end{align}
where $d\Omega^2=d\theta^2+\sin^2\theta d\varphi^2$ is the standard line element on a unit two-sphere {and,} $\Phi (r)$ and $b(r)$ are denoted as redshift and shape functions, respectively. The radial coordinate ranges from $r_0$ to spatial infinity where the surface at $r=r_0$ is known as the wormhole\rq{}s throat. Although the radial metric component {\bf i.e.,} $g_{rr}$ diverges in the limit $r\rightarrow r_0$, the proper radial distance~\cite{mt,FLoboBook}
\be\label{properl}
l(r)=\pm\int_{r_0}^{r}\f{dr}{\sqrt{1-\f{b(r)}{r}}},
\ee
is required to be finite everywhere. This quantity decreases from $\ell=+\infty$ in the upper universe to $\ell=0$ at the throat, and then from zero to $\ell=-\infty$ in the lower universe. We now define the time-like and space-like vector fields, respectively as $u^\mu=[{\rm e}^{-\Phi(r)},0,0,0]$ and $v^\alpha=\left[0,\sqrt{1-b(r)/r},0,0\right]$, so that $u^\mu u_\mu=-1$ and $v^\alpha v_\alpha=1$. The anisotropic {\rm EMT} of matter source then takes the form
\be\label{emtaniso}
{\rm T}_{\mu\nu}=[\rho(r)+p_t(r)]u_\mu u_\nu+p_t(r)g_{\mu\nu}+[p_r(r)-p_t(r)]v_\mu v_\nu,
\ee
with $\rho(r)$, $p_r(r)$, and $p_t(r)$ being the energy density, radial and tangential pressures, respectively. We therefore obtain the components of field equation (\ref{bdfield}) as 
\bea
16\pi\rho(r)&=&\f{2\phi b^\prime}{r^2}-2\left(1-\f{b}{r}\right)\phi^{\prime\prime}-\f{\omega}{\phi}\left(1-\f{b}{r}\right)(\phi^\prime)^2+\f{\phi^\prime}{r}\left(b^\prime+3\f{b}{r}-4\right)-V(\phi),\label{rhoex}\\
16\pi p_r(r)&=&-\f{\omega}{\phi}\left(1-\f{b}{r}\right)(\phi^\prime)^2+2\left(1-\f{b}{r}\right)\left(\Phi^\prime+\f{2}{r}\right)\phi^\prime+4\phi\left(1-\f{b}{r}\right)\f{\Phi^\prime}{r}-2\f{\phi b}{r^3}+V(\phi),\label{prexp}\\
16\pi p_t(r)&=&2\left(1-\f{b}{r}\right)\Phi^{\prime\prime}+2\phi\left(1-\f{b}{r}\right)\Phi^{\prime\prime}+\f{\omega}{\phi}\left(1-\f{b}{r}\right)(\phi^\prime)^2+\f{2}{r}\left[(r-b)\Phi^\prime-\f{b^\prime}{2}+1-\f{b}{2r}\right]\phi^\prime\nn&+&2\phi\left(1-\f{b}{r}\right)(\Phi^\prime)^2-\f{\phi}{r}\left(b^\prime-2+\f{b}{r}\right)\Phi^\prime-\f{b^\prime}{2r^2}+\f{b}{2r^3}+\f{V(\phi)}{2\phi},\label{ptexp}
\eea
where a prime denotes differentiation with respect to radial coordinate $r$. Also the evolution equation (\ref{bdevol}) reads
\be\label{bdevol1}
\left(1-\f{b}{r}\right)\phi^{\prime\prime}+\left[(r-b)\Phi^\prime-\f{b^\prime}{2}-\f{3b}{2r}+2\right]\f{\phi^\prime}{r}+\f{1}{2\omega+3}\left[8\pi(\rho-p_r-2p_t)+2V-\phi\f{dV}{d\phi}\right]=0.
\ee
The conservation of {\rm EMT}, i.e., $\nabla^\mu T^{(m)}_{\mu\nu}=0$ leads to the following equation
\be\label{conseq}
\Phi^\prime(\rho+p_r)+p_r^\prime+\f{2}{r}\left(p_r-p_t\right)=0.
\ee
In view of the above system of differential equations we find that on substitution for energy density and pressure profiles from Eqs.~(\ref{rhoex})-(\ref{ptexp}) into equation (\ref{conseq}) and after simplification we arrive at evolution equation for BD scalar field, i.e., Eq.~(\ref{bdevol1}). This means that, from equations (\ref{rhoex})-(\ref{conseq}) only four of them are independent. Having this in mind, we proceed to find wormhole configurations assuming matter energy density, Eq.~(\ref{rhoex}), obeys the Casimir energy density for a specified {\bf configuration}. We therefore set
\be\label{rhoehoc}
\rho(r)=\rho_c(r).
\ee
Substituting for energy density and pressure profiles into Eq.~(\ref{bdevol1}) gives
\bea\label{bdevol2}
&&2r\phi(2\omega+1)(r-b)\phi^{\prime\prime}-r\omega(r-b)(\phi^\prime)^2-\phi\phi^\prime\left[(rb^\prime-4r+3b)(2\omega+1)-4r\omega(r-b)\Phi^\prime\right]\nn&-&4r\phi^2(r-b)\Phi^{\prime\prime}+2\phi\left[\phi\Phi^\prime(rb^\prime-4r+3b)+\phi b^\prime-2r\phi(r-b)(\Phi^\prime)^2-r^2\left(\f{\phi}{\phi^\prime}V^\prime-\f{V}{2}\right)\right]\nn&+&16\pi\lambda\phi r^{2-m}=0.
\eea
Equations (\ref{rhoehoc}) and (\ref{bdevol2}) construct a system of two coupled differential equations to be solved for four unknowns $\{b(r),\phi(r),\Phi(r),V(r)\}$. Obviously this system is under-determined, hence, in order to solve it we have to specify the functionality of two of the unknowns. This is our task in the next section in which we assume physically reasonable {forms for} the BD scalar field and the redshift function and {try to find the} scalar field potential and the shape function. 
\subsection{Solutions with $\Phi(r)=0$}\label{redzero}
{Our attempt in the present subsection is to build and study a class of zero tidal force wormhole solutions whose energy density is given by Eq.~(\ref{rhoehoc})}. We firstly note that there should be no horizon in wormhole spacetime, since the presence of horizon which is defined as the surface with ${\rm e }^{2\Phi(r)}\rightarrow0$ would prevent two-way travel through the wormhole; hence, the redshift function must be finite everywhere so that there is no singularity and event horizon in the wormhole spacetime. In order that the shape function represents a physically reasonable wormhole solution it has to fulfill the fundamental condition $b(r_0)=r_0$ along with the flare-out condition, i.e., $rb^\prime-b<0$ throughout the spacetime and also at the throat; therefore at the throat we must have $b^\prime(r_0)<1$~\cite{mt,mt1}. Also, for $r>r_0$ the radial metric component $g_{rr}=\left(1-b(r)/r\right)^{-1}$ must be positive or the shape function must satisfy the condition $1-b(r)/r>0$. This condition guarantees that the (Lorentzian) metric signature is preserved for radii bigger than the throat radius. Let us now proceed to find the shape function and scalar field potential, assuming the functionality of BD scalar field as $\phi(r)=\phi_0\left(\f{r_0}{r}\right)^n$ where $n\in\mathbb{R}^+$ and $\phi_0$ is the value of scalar field at the throat. The system (\ref{rhoehoc}) and (\ref{bdevol2}) then admits the following exact solution for $n=2$ and $m\neq2$, given as
\bea
b(r)\!\!\!\!&=&\!\!\!\!\f{4(\omega+1)r}{4\omega+5}+\f{\left[(m-2)\phi_0-8\pi\lambda m r_0^{2-m}\right]{r}^{3}}{\left(4\omega+5\right)(m-2)r_0^{2}\phi_0}+\f{8\pi\lambda mr^{5-m}}{
r_0^{2}(m-2)(4\omega+5)\phi_0},\label{solb}\\
V(r)\!\!\!\!&=&\!\!\!\!\f{4\phi_0r_0^2(1+\omega)}{(4\omega+5)r^4}-\f{32\pi\lambda[m(1+\omega)-4\omega-5]}{(m-2)(4\omega+5)r^m}-\f{(2\omega+3)\left(16\pi\lambda mr_0^{2-m}-2(m-2)\phi_0\right)}{(m-2)(4\omega+5)r^2},\label{solV}
\eea
where the integration constant has been set according to the condition $b(r_0)=r_0$. From Eq.~(\ref{solb}) the flare-out condition at the throat leads to the following inequality
\bea\label{flr0}
\f{\phi_0(4\omega+7)-8\pi\lambda mr_0^{2-m}}{(4\omega+5)\phi_0}<1.
\eea
In the framework of {\rm GR}, the fundamental flaring-out condition entails the violation of {\rm NEC}. In fact, traversability of wormhole spacetimes put constraints on the EMT which must be fulfilled at (or near) any wormhole throat. These constraints follow from combining the null focusing theorem, through the Raychaudhuri equation, with the flare-out condition along with using the Einstein field equation~\cite{khu1}. In a traversable wormhole spacetime, one may apply the focusing theorem on a congruence of null rays defined by {the} null vector field $k^\mu$ with $k^\mu k_\mu=0$~\cite{PoissonBook}. The traversability of wormhole then requires ${\rm T}_{\mu\nu}k^\mu k^\nu\leq0$, which is nothing but the violation of NEC by the EMT of an exotic matter~\cite{Hochberg1998}. For the {\rm EMT} given in (\ref{emtaniso}) the {\rm NEC} is given by 
\bea
&&\rho(r)+p_{r}(r)\geq0,~~~~~~~~\rho(r)+p_{t}(r)\geq0\label{nec}.
\eea
Also, for the sake of physical validity of the solutions, we require that the wormhole configuration respects the WEC. This condition asserts that, any observer with {the} timelike 4-vector $u^\mu$ in the spacetime must measure non-negative values for the energy density or ${\rm T}_{\mu\nu}u^\mu u^\nu\geq0$~\cite{PoissonBook}. For the EMT given in (\ref{emtaniso}), {\rm WEC} leads to the following inequalities
\bea
&&\rho(r)\geq0,~~~~~\rho(r)+p_{r}(r)\geq0,~~~~~\rho(r)+p_{t}(r)\geq0.\label{wec}
\eea
We note that {\rm WEC} implies the null form. For the solutions obtained in Eqs.~(\ref{solb}) and (\ref{solV}) we {find} the EMT components as
\bea
&&\rho(r)=\f{\lambda}{r^m},\label{energy}\\
&&p_{r}(r)=\f{(3+2\omega)(\phi_0(m-2)-8\pi\lambda mr_0^{2-m})}{4\pi(4\omega+5)(m-2)r^2}+\f{\lambda(m+8\omega+10)}{(4\omega+5)(m-2)r^m}-\f{\phi_0r_0^2(2\omega+3)}{4\pi(4\omega+5)r^4},\label{pressr}\\
&&p_{t}(r)=\f{\phi_0r_0^2(2\omega+3)}{4\pi(4\omega+5)r^4}-\f{\lambda(m+8\omega+10)}{2(4\omega+5)r^m},\label{presst}
\eea
whence we get
\bea
\rho(r)+p_{r}(r)\!\!\!\!&=&\!\!\!\!\f{(2\omega+3)\left(\phi_0(m-2)-8\pi\lambda mr_0^{2-m}\right)}{4\pi(4\omega+5)(m-2)r^2}+\f{(2\omega+3)\left[8\pi\lambda mr^{4-m}-\phi_0r_0^2(m-2)\right]}{4\pi(4\omega+5)(m-2)r^4},\label{energypr}\\
\rho(r)+p_{t}(r)\!\!\!\!&=&\!\!\!\!\f{(2\omega+3)\phi_0r_0^2}{4\pi(4\omega+5)r^4}-\f{\lambda m}{2(4\omega+5)r^{m}}.\label{energypt}
\eea
Hence, the energy conditions at the throat take the form
\bea
\rho\Big|_{r=r_0}\!\!\!\!\!&=&\!\!\!\!\!\f{\lambda}{r_0^m}\geq0,~~~~~~~~\rho+p_{r}\Big|_{r=r_0}=0,~~~~~~~\rho+p_{t}\Big|_{r=r_0}=\f{(2\omega+3)\phi_0-2\pi\lambda mr_0^{2-m}}{4\pi(4\omega+5)r_0^2}\geq0.\label{enprpt0}
\eea
From the first part of Eq.~(\ref{enprpt0}) we observe that the positivity of energy density at wormhole throat requires that $\lambda>0$. This leads to a repulsive Casimir force. From the second part we find that at the throat, the matter distribution in radial direction acts as dark energy, assuming a linear equation of state (EoS) $p_r(r_0)=-\rho(r_0)$. Such a situation can also occur in tangential direction if we take the BD coupling parameter as
\be\label{phi0rhopt}
\omega=\f{\pi\lambda m}{\phi_0r_0^{m-2}}-\f{3}{2}.
\ee
In this regard, in the limit of approach to the throat, one may recognize the matter distribution as a ring of dark energy, either in radial or tangential directions. From the third part of Eq.~(\ref{enprpt0}) we find that the fulfillment of NEC in tangential direction depends on values of {the parameters} $\{\phi_0,\omega,\lambda,r_0,m\}$. These construct a five dimensional space parameter that the allowed values of which determine physically reasonable wormhole configurations. The two parameters $\{\lambda,m\}$ can be specified through the geometry and shape of the objects for which the Casimir effect is considered. In the present study we examine the satisfaction of conditions on wormhole solutions assuming $m=4$. For this case the corresponding Casimir energy density of a scalar field in a configuration of two infinitely extended parallel plates can be obtained by setting $\lambda=-\pi^2/1440$ for Dirichlet boundary condition and $\lambda=(7/8)\times\pi^2/1440$ for mixed (Dirichlet on one plane {and} Neumann on the other {plane}) boundary condition\footnote{The scalar Casimir effect with Robin boundary conditions on two parallel plates has been considered in~\cite{Robinbound}}~\cite{advances,milton2003sf,actor1996,milton2004}. As the Lorentzian signature must be preserved for $r>r_0$ we need to check  this condition for the shape function (\ref{solb}). {For $m=4$ the} roots of inverse of radial metric component at the throat, i.e., $g_{rr}^{-1}(r_0)=1-b(r_0)/r_0$, are given by $r_1=r_0$ and $r_2=-r_0$. Clearly the second root is not a physical value for the throat radius. Also, the proper distance as a function of radial coordinate is found as
\be\label{propl}
l(r)=\alpha r_0^2\left(\f{(4\omega+5)\phi_0}{16\pi\lambda-r_0^2\phi_0}\right)^{\f{1}{2}}\ln\left[\f{r_0}{r+\sqrt{r^2-r_0^2}}\right],
\ee
where $\alpha=-1,+1$ determine the upper and lower universes, respectively and $l(r_0)=0$. Now, {one can realize that} the flare-out condition (\ref{flr0}), the third part of Eq.~(\ref{enprpt0}) and the reality condition on proper distance restrict the values of the parameters $\{\phi_0,\omega,r_0\}$ according to the following inequalities
\bea\label{ineqs4}
&&\!\!\!\!\!\!\!\!\Bigg\{\phi_0<0\land\Bigg[\left(0<r_0<\f{\pi^{\f{3}{2}}}{3\sqrt{-10\phi_0}}\land-\f{5}{4}<\omega\leq-\f{\pi^3}{360r_0^2\phi_0}-\f{3}{2}\right)\nn&\lor&\!\!\!\!\!\!\left(r_0>\f{\pi^{\f{3}{2}}}{3\sqrt{-10\phi_0}}\land-\f{\pi^3}{360r_0^2\phi_0}-\f{3}{2}\leq\omega<-\f{5}{4}\right)\Bigg]\Bigg\}\!\lor\!\Bigg\{\phi_0>0\land r_0>0\land\omega\leq-\f{\pi^3}{360r_0^2\phi_0}-\f{3}{2}\Bigg\},\nn
\eea
\begin{figure}
	\begin{center} 
		\includegraphics[width=7.7cm]{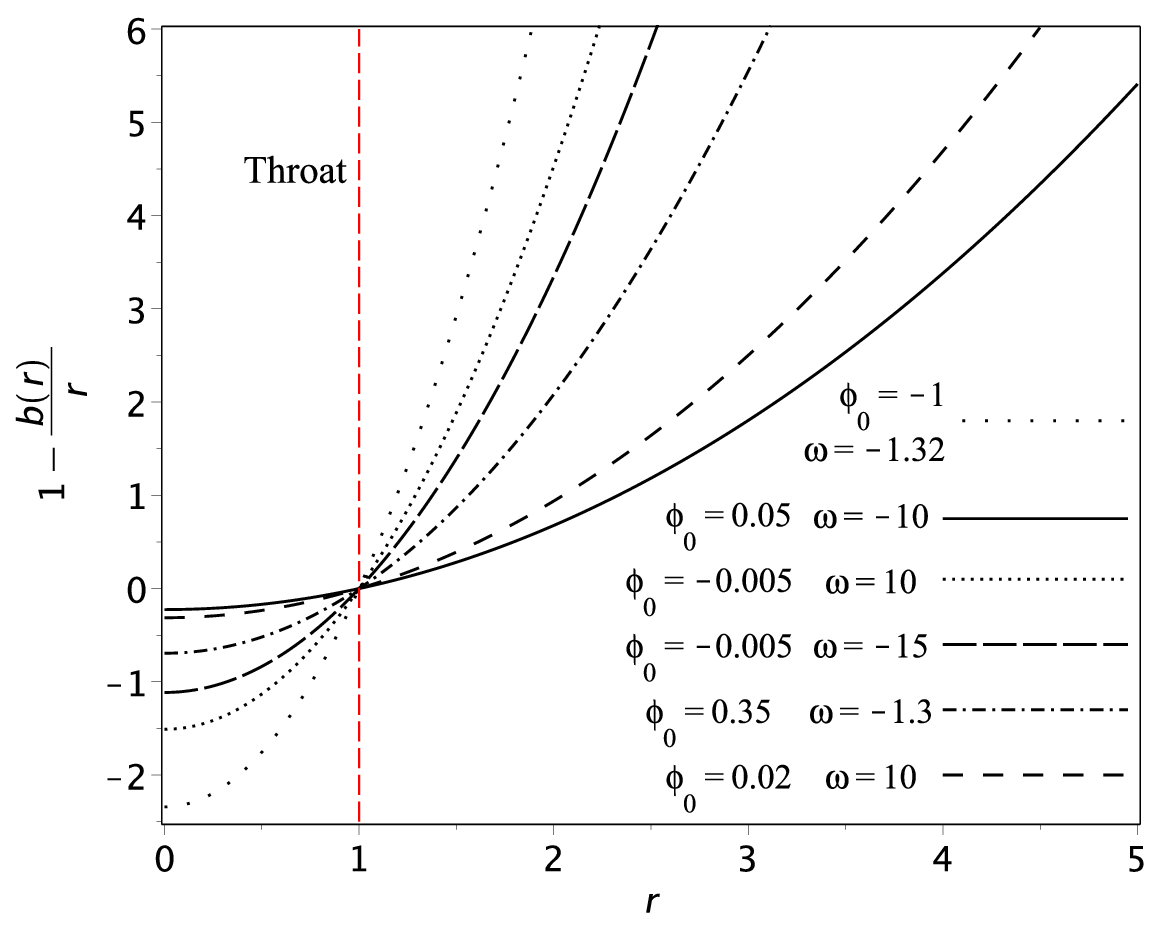}
		\includegraphics[width=7.7cm]{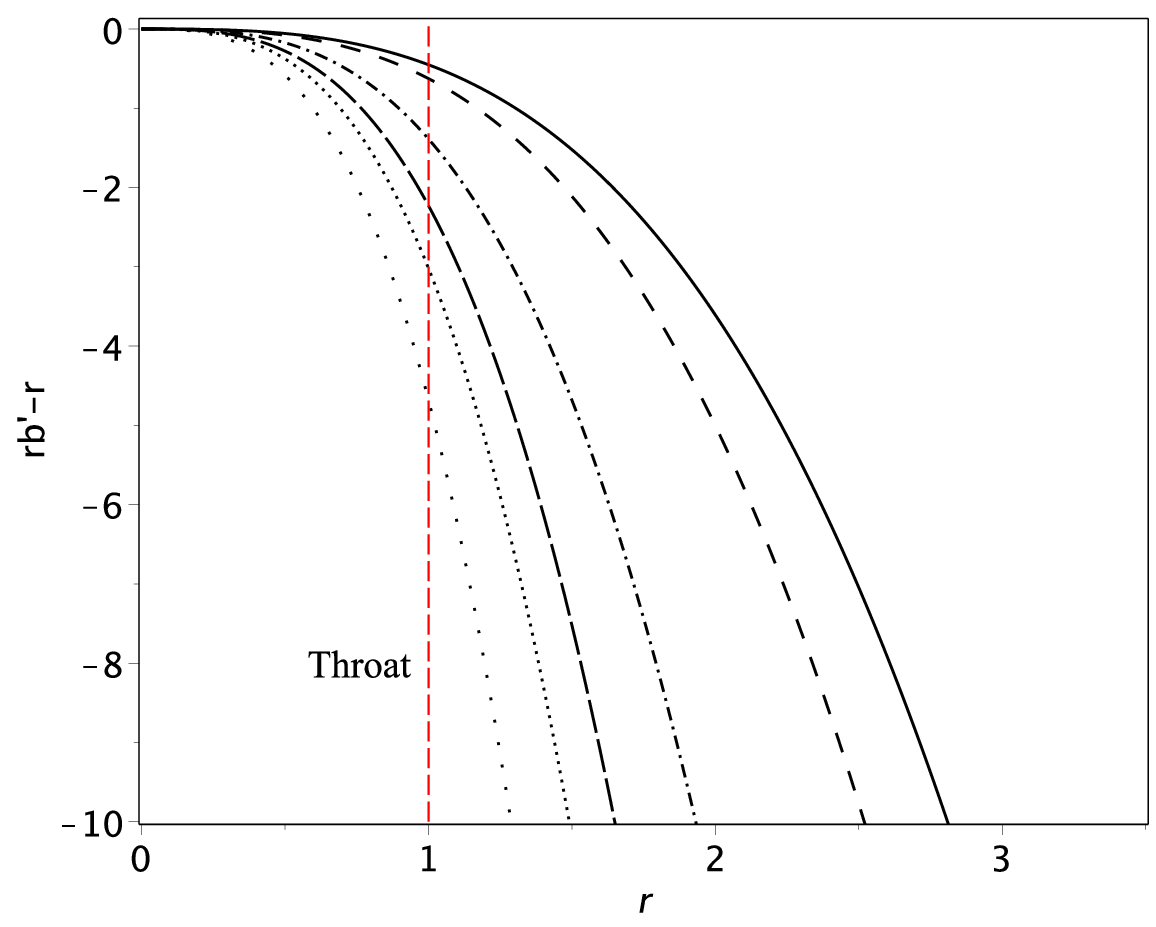}
		\caption{Left panel: Behavior of $g_{rr}^{-1}$ for $m=4$, $r_0=1$. The solid, dotted and space-dotted curves have been plotted for $\lambda=-\pi^2/1440$. The dot-dashed, space-dashed and long-dashed curves have been plotted for $\lambda=(7/8)\times\pi^2/1440$. Right panel: Behavior of flare-out condition for the same values of model parameters as of the left panel with the same line styles.}\label{fig1}
	\end{center}
\end{figure}
for $\lambda=-\pi^2/1440$, and 
\bea\label{ineqs14}
&&\!\!\!\!\!\!\!\!\Bigg\{\phi_0>0\land\Bigg[\left(0<r_0<\f{\pi}{12}\sqrt{\f{7\pi}{5\phi_0}}\land\omega\geq\f{7\pi^3}{2880r_0^2\phi_0}-\f{3}{2}\right)\nn&\lor&\!\!\!\!\!\!\left(r_0>\f{\pi}{12}\sqrt{\f{7\pi}{5\phi_0}}\land\omega\leq\f{7\pi^3}{2880r_0^2\phi_0}-\f{3}{2}\right)\Bigg]\Bigg\}\!\lor\!\Bigg\{\phi_0<0\land r_0>0\land\f{7\pi^3}{2880r_0^2\phi_0}-\f{3}{2}\leq\omega<-\f{5}{4}\Bigg\},\nn
\eea
for $\lambda=(7/8)\times\pi^2/1440$. From the above expressions we observe that spatial extension of wormhole and allowed values of BD parameter depend on whether the Casimir force be attractive or repulsive. We further deduce that the upper and lower limiting values of BD coupling parameter in the limit $\omega\rightarrow-\f{3}{2}^{\pm}$ can occur for two cases: $i)$ setting $\phi_0<\pm \infty$ and $r_0\rightarrow\infty$ and, $ii)$ $r_0<\infty$ and $\phi_0\rightarrow\pm\infty$. The former case is not reasonable as the throat radius must be finite. The latter signals a singular behavior for the scalar field at the throat which is physically untenable. Hence, in the limit $\omega\rightarrow-\f{3}{2}^{\pm}$, the model parameters may not provide physically reasonable wormhole solutions. Moreover, we note that the obtained bounds on BD parameter can have overlaps with those reported in the previous works. For example, static spherically symmetric wormhole solutions have been obtained in vacuum BD theory with coupling parameter lying in the interval\footnote{Due to singular behavior of the solutions, this interval for BD parameter was later revised to $-2<\omega<-4/3$~\cite{bd3}.} $-3/2<\omega<-4/3$~\cite{bd3rev}; depending on the values of $\{r_0,\phi_0\}$, this interval for BD coupling parameter can lay within the ranges given in the second lines of Eqs.~(\ref{ineqs4}) and (\ref{ineqs14}) for $\phi_0<0$. Also, in~\cite{bd6} a wormhole analogy to Horowitz-Ross naked black holes~\cite{HorRonb} has been developed for $\omega<-2$. This upper bound on BD coupling parameter corresponds to the second line of Eq.~(\ref{ineqs4}) for $\phi_0>0$. By the same logic as in our analysis, the bounds obtained in~\cite{bd4} as $\omega<-2$ or $-2<\omega\leq0$ can be matched to the ranges given in Eqs.~(\ref{ineqs4}) and (\ref{ineqs14}) through suitable choice of the throat radius and value of BD scalar field at the throat. 
\par
\begin{figure}
	\begin{center}
		\includegraphics[width=7.7cm]{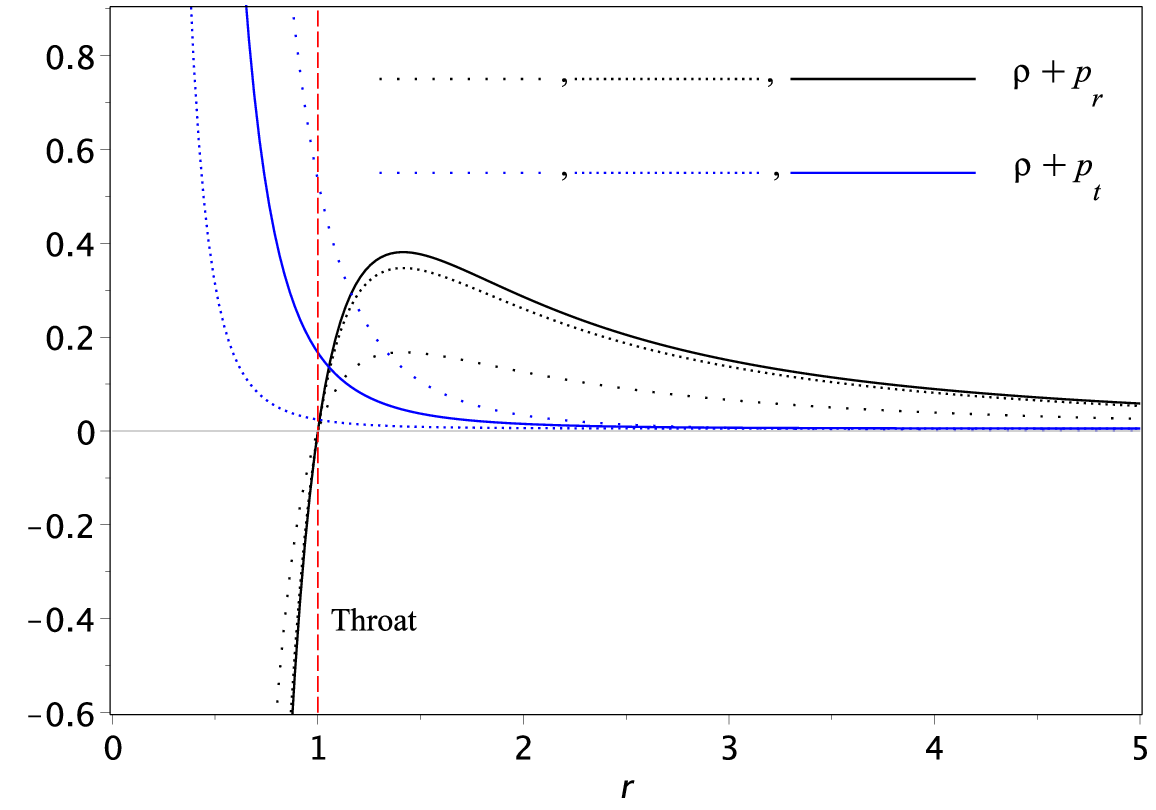}
		\includegraphics[width=7.7cm]{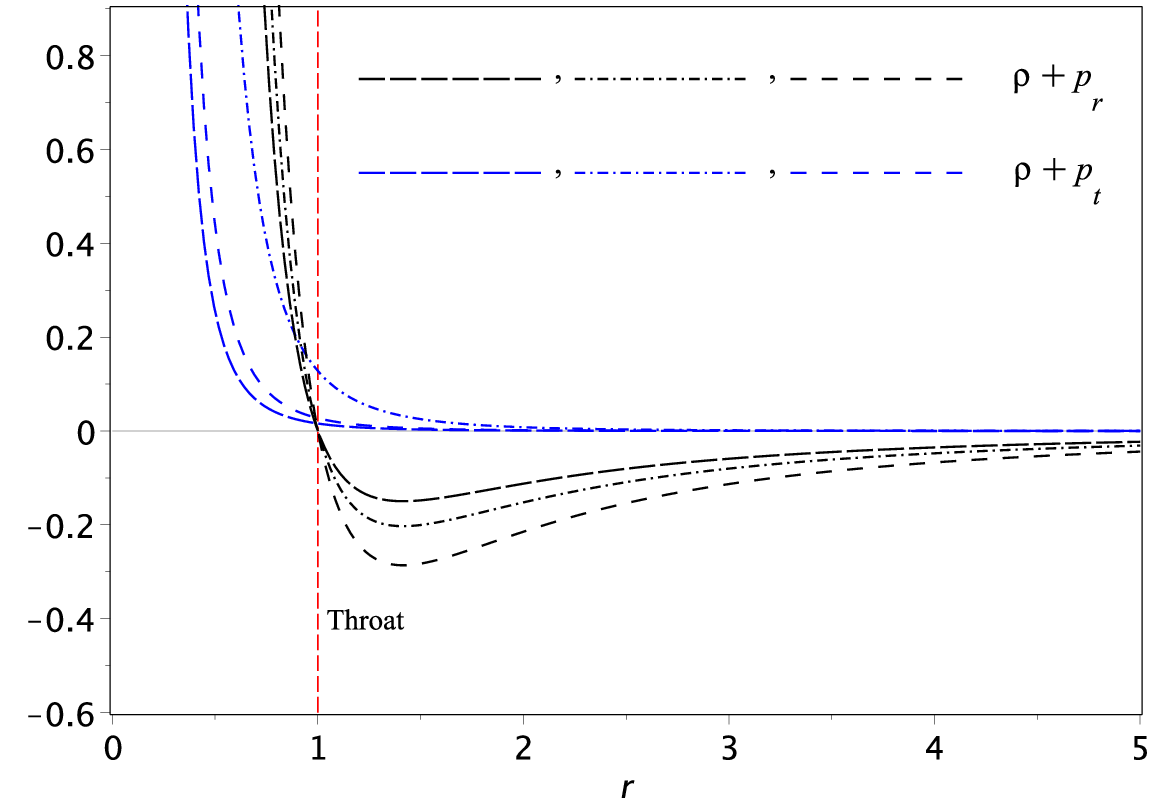}
		\caption{Behavior of NEC for $m=4$ and $r_0=1$, $\lambda=-\pi^2/1440$ (left panel) and $\lambda=(7/8)\times\pi^2/1440$ (right panel). The other parameters have been set as those of left panel in Fig.~(\ref{fig1}) with the same line styles.}\label{fig2}
	\end{center}
\end{figure}
\begin{figure}
	\begin{center}
		\includegraphics[width=7.3cm]{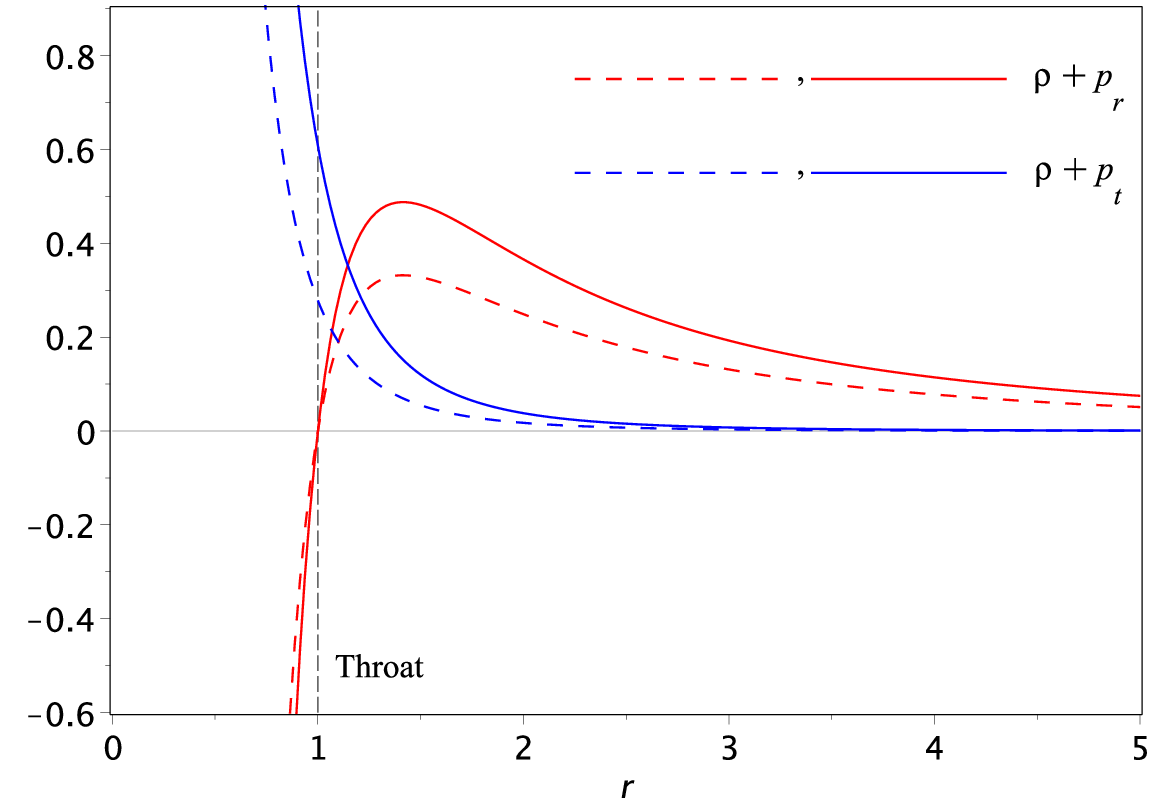} 
		\includegraphics[width=8.2cm]{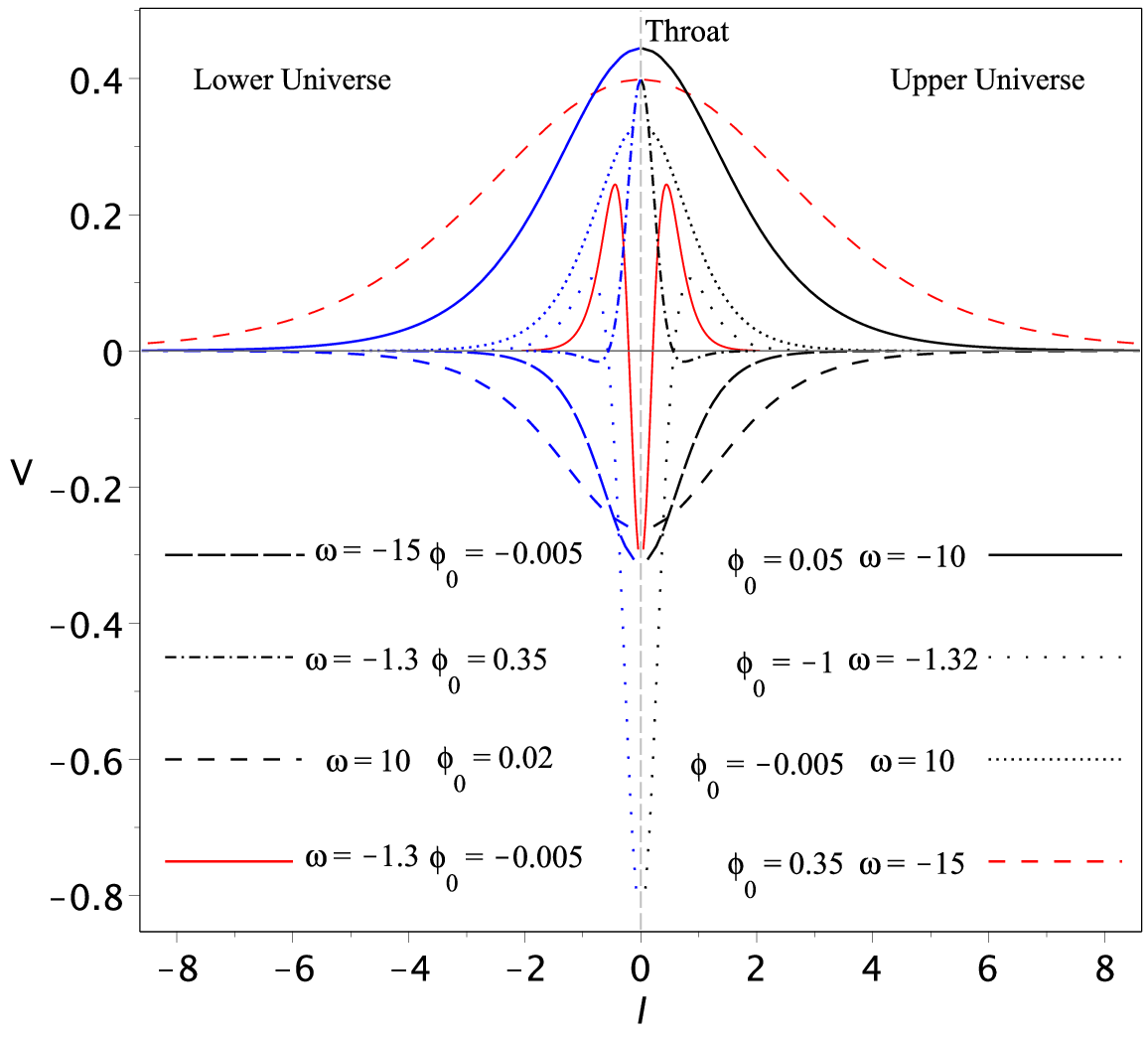}
		\caption{Left panel: Behavior of NEC for $m=4$ and $r_0=1$, $\lambda=(7/8)\times\pi^2/1440$, $\phi_0=-0.005$ and $\omega=-1.3$ (solid curves) and, $\phi_0=0.35$ and $\omega=-15$ (dashed curves). Right panel: Behavior of scalar field potential against the proper distance. The model parameters have been set as of the left panel in Fig.~(\ref{fig1}) with the same line styles. The red curves have been plotted for $\lambda>0$.}\label{fig3}
	\end{center}
\end{figure}

The left panel in Fig.~(\ref{fig1}) presents the inverse of radial metric component where we observe that Lorentzian signature is preserved throughout the spacetime. The values of parameters $\{r_0,\omega,\phi_0\}$ have been chosen according to the inequalities (\ref{ineqs4}) for $\lambda<0$ and (\ref{ineqs14}) for $\lambda>0$. To be a solution of a wormhole, we need to impose that the throat flares out. This condition can be checked through the behavior of $rb^\prime(r)-b(r)$ which has been sketched in the right panel. It is therefore seen that this quantity is negative for $r>r_0$, hence, the shape function satisfies the flare-out condition. The left panel in Fig.~(\ref{fig2}) shows the behavior of NEC in radial (family of black curves) and tangential (family of blue curves) directions for $\lambda<0$, where we observe that the NEC is satisfied at the throat and beyond it. We note that the values of parameters $\{\phi_0,\omega\}$ have been chosen according to inequalities (\ref{ineqs4}) estimated at $r_0=1$,
\bea\label{ineqs4r0}
&&\left\{\phi_0<-\f{\pi^3}{90}\land-\f{\pi^3}{360\phi_0}-\f{3}{2}\leq\omega<-\f{5}{4}\right\}\lor\left\{-\f{\pi^3}{90}<\phi_0<0\land-\f{5}{4}<\omega\leq-\f{\pi^3}{360\phi_0}-\f{3}{2}\right\}\nn&&\lor\left\{\phi_0>0\land\omega\leq-\f{\pi^3}{360\phi_0}-\f{3}{2}\right\}.
\eea
Also the NEC is satisfied in tangential direction throughout the spacetime for $\lambda>0$, see the family blue curves in the right panel of Fig.~(\ref{fig2}). The radial profile of NEC is fulfilled at the throat but it is violated for $r>r_0$, see the family of black curves. However, we note that since $\rho(r_0)+p_r(r_0)=0$, then the quantity $\rho(r)+p_r(r)$ can change its sign at the throat provided that the slope of the curve is
positive at $r=r_0$. In other words for $r_0=1$ we have
\be\label{slopenecr}
\f{d}{dr}\left[\rho(r)+p_r(r)\right]\Big|_{r=r_0}\!\!\!\!\!\!>0\Rightarrow\left\{\phi_0<\f{7\pi^3}{720}\land\f{-3}{2}<\omega<\f{-5}{4}\right\}\lor\left\{\phi_0>\f{7\pi^3}{720}\land\left(\omega<\f{-3}{2}\lor\omega>\f{-5}{4}\right)\right\}.
\ee
The above conditions along with the inequalities (\ref{ineqs14}) estimated at $r_0=1$ further restrict the values of parameters $\{\phi_0,\omega\}$ at the throat as
\be\label{finalranecr}
\left\{\phi_0<0\land-\f{3}{2}<\omega<-\f{5}{4}\right\}\lor\left\{\phi_0>\f{7\pi^3}{720}\land\omega<-\f{3}{2}\right\}.
\ee
Hence, the NEC in radial direction can be satisfied at the throat and throughout the spacetime if we choose the values of parameters $\{\phi_0,\omega\}$ subject to the inequalities given in (\ref{finalranecr}), see the red and blue curves in the left panel of Fig.~(\ref{fig3}). We therefore conclude that zero tidal force wormhole solutions in BD theory with Casimir energy as the matter source can be built without violating NEC and WEC (for $\lambda>0$). In the right panel of Fig.~(\ref{fig3}) the behavior of scalar field potential against the proper distance is sketched where it is seen that in the limit of approach to the throat, the BD scalar field faces a potential barrier at the throat where $l=0$. The height of this barrier depends on $\{\lambda,\omega,\phi_0\}$ parameters, see the solid (black and blue) and dotted  curves for $\lambda<0$ and, dot-dashed and space-dashed (red) curves for $\lambda>0$. Moreover, at the throat, the potential can assume maximum values in negative direction that the depth of this potential wells depends on model parameters, see space-dotted, space-dashed (black and blue) and long-dashed curves. Also, in the regions near the throat ($l>0$ in the upper universe and $l<0$ in the lower universe) where the potential changes its sign we can have potential barrier (solid red and space-dotted curves) and potential well (dot-dashed curve) that the height and depth of which depend on the values of model parameters. Finally, let us take the value of BD scalar field at the throat as
\be\label{morth}
\phi_0=\f{8\pi\lambda mr_0^{2-m}}{m-2},~~~~~m\neq\{2,4\},~~~~~\omega\neq-\f{5}{4},
\ee
then from Eq.~(\ref{solb}) we get 
\be\label{bsolmt}
1-\f{b(r)}{r}=\f{1}{4\omega+5}\left[1-\left(\f{r_0}{r}\right)^{m-4}\right].
\ee
Now, if we consider the Casimir effect for a configuration of objects in such a way that $m=6$, see e.g.,~\cite{m6cas}, then the line element (\ref{metric}) assumes the following form
\begin{align}\label{metricmt}
ds^2=-dt^2+\f{d\tilde{r}^2}{1-\left(\f{\tilde{r}_0}{\tilde{r}}\right)^2}+\f{\tilde{r}^2}{(4\omega+5)}d\Omega^2,
\end{align}
where use has been made of the rescaling $r^2=\tilde{r}^2/(4\omega+5)$. The above metric represents the well-known {MT} spacetime~\cite{mt} with a solid angle deficit for $\omega>-1$ and solid angle excess for $-5/4<\omega<-1$. We note that the solid angle of a unit two-sphere is now $4\pi/(4\omega+5)<4\pi$ for $\omega>-1$ and $4\pi/(4\omega+5)>4\pi$ for $-5/4<\omega<-1$. For the special case with $\omega=-1$, which corresponds to the low energy limit of some string theories~\cite{omm1ls},\cite{Faraoni}, the above solution reduces to the standard {MT} metric.
\subsection{Solutions with $\Phi(r)\neq0$}\label{nonzerored}
\begin{figure}
		\begin{center}
			\includegraphics[width=7.0cm]{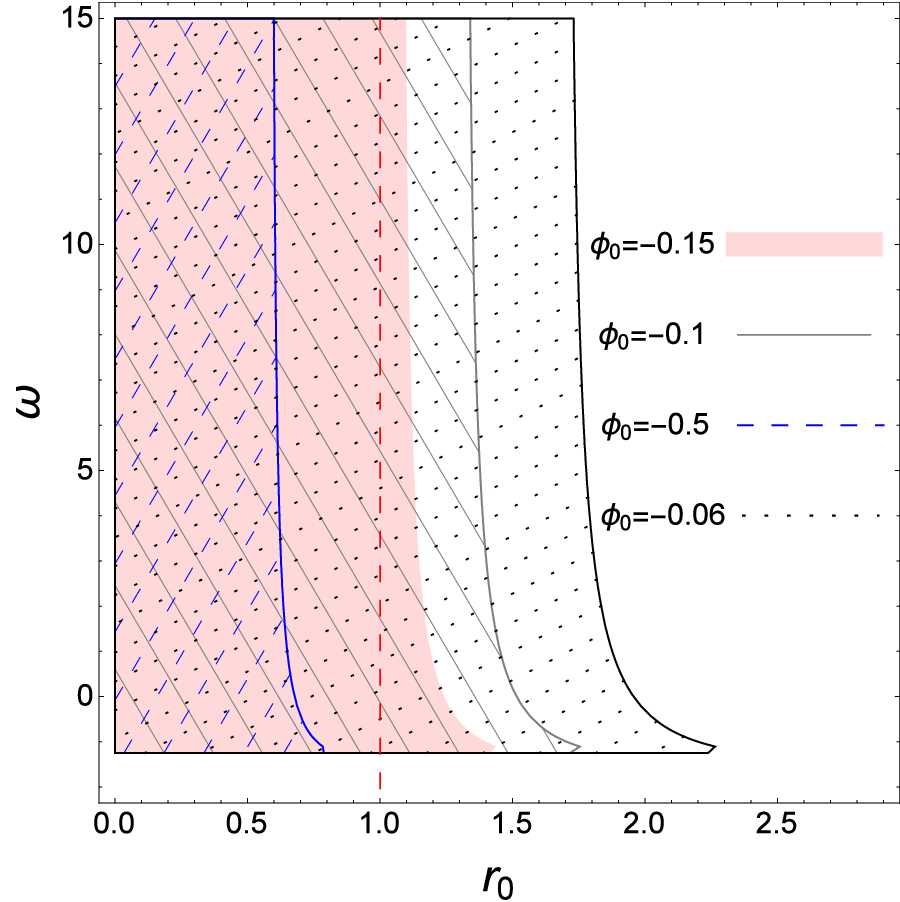}
			\includegraphics[width=7.7cm]{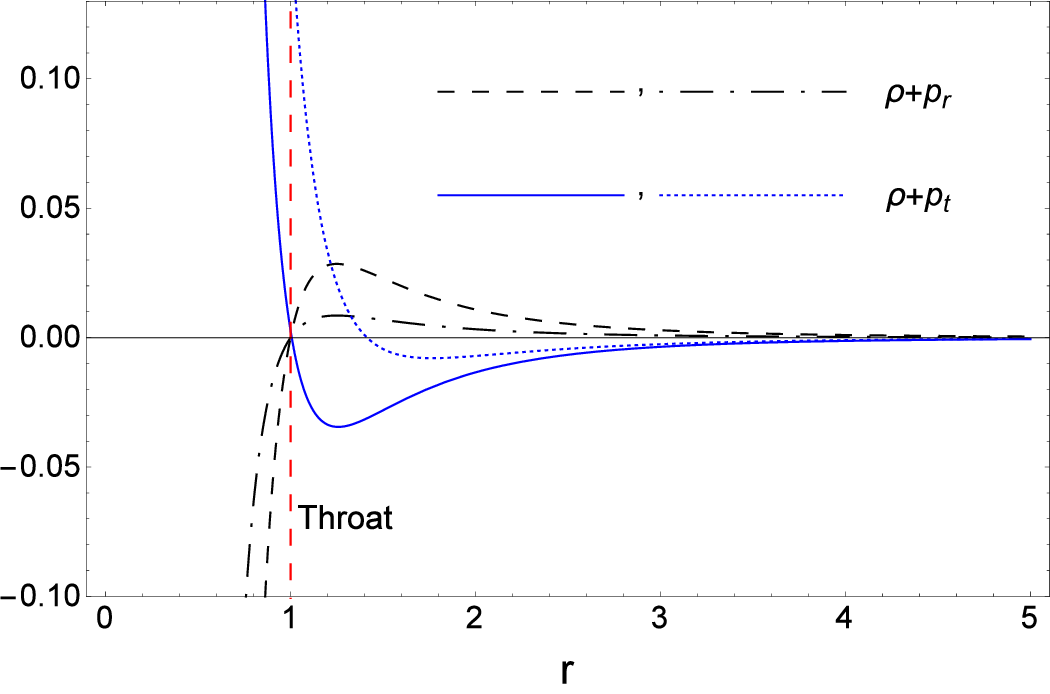}
			\caption{Left panel: The allowed regions for BD coupling parameter and throat radius for $\lambda=-\pi^2/1440$, and different negative values of $\phi_0$. Right panel: The behavior of NEC in radial (black curves) and tangential (blue curves) directions for $r_0=1$ and $\lambda=-\pi^2/1440$. The model parameters have been chosen from the left panel, as: $\phi_0=-0.15$, $\omega=-1.20$ (dashed and dotted curves) and $\omega=10$ (solid and dot-dashed curves).}\label{fig4}
		\end{center}
	\end{figure}

\begin{figure}
	\begin{center}
		\includegraphics[width=7.7cm]{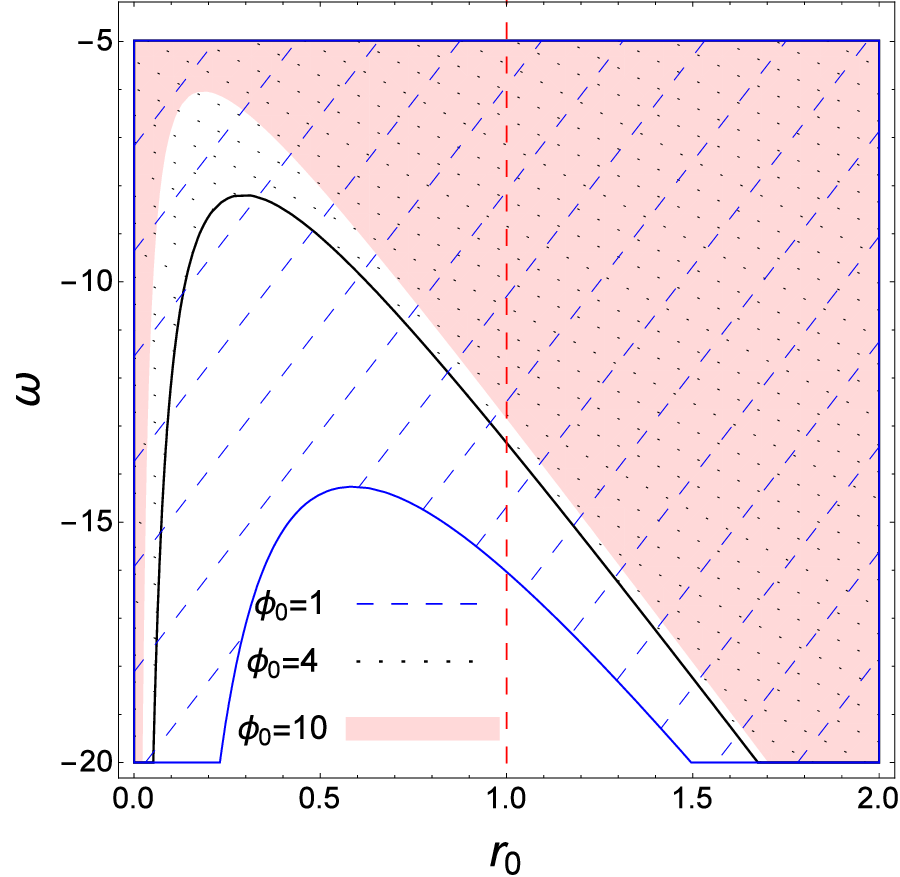}
		\includegraphics[width=7.7cm]{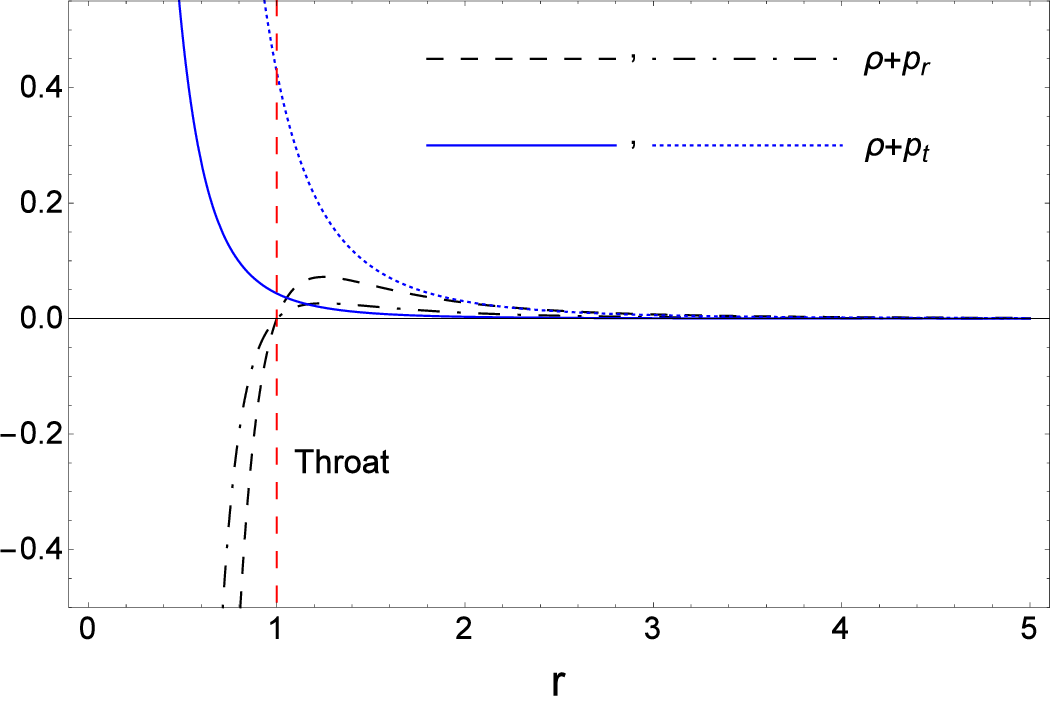}
		\caption{Left panel: The allowed regions for BD coupling parameter and throat radius for $\lambda=-\pi^2/1440$, and different positive values of $\phi_0$. Right panel: The behavior of NEC in radial (black curves) and tangential (blue curves) for $r_0=1$ and $\lambda=-\pi^2/1440$. The model parameters have been set as $\omega=-10$ and $\phi_0=10$ (dashed and dotted curves) and $\phi_0=1$ (solid and dot-dashed curves).}\label{fig5}
	\end{center}
\end{figure}

\begin{figure}
	\begin{center}
		\includegraphics[width=7.7cm]{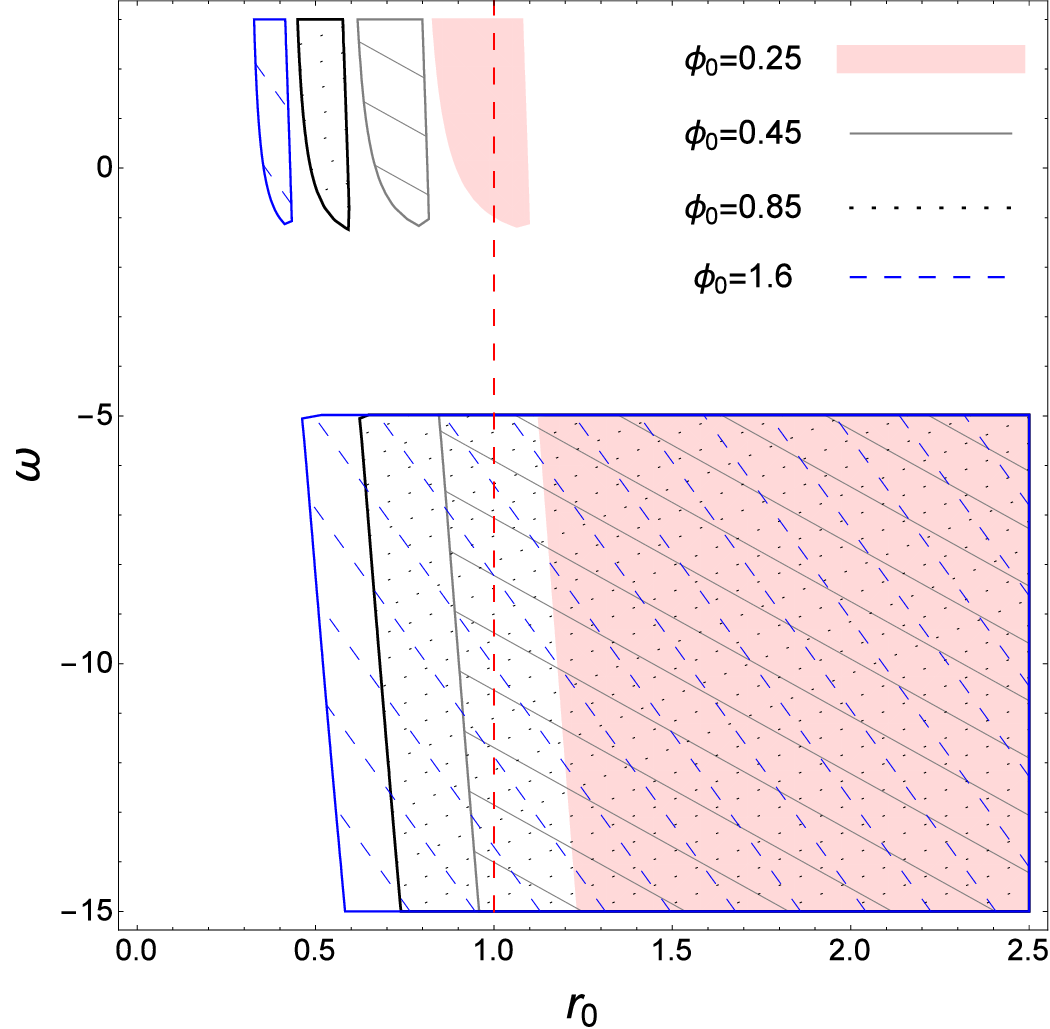}
		\includegraphics[width=7.7cm]{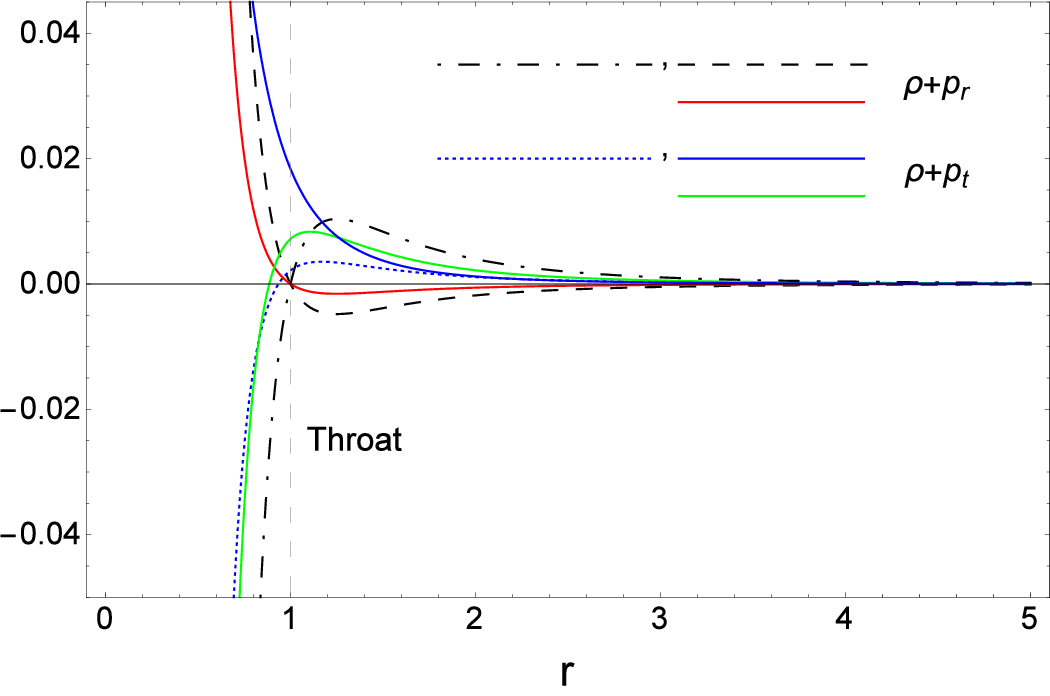}
		\caption{Left panel: The allowed regions for BD coupling parameter and throat radius for $\lambda=(7/8)\times\pi^2/1440$, and different values of $\phi_0$. Right panel: The behavior of NEC in radial (black and red curves) and tangential (blue and green curves) for $r_0=1$ and $\lambda=(7/8)\times\pi^2/1440$. The model parameters have been set as $\phi_0=0.25$, $\omega=-0.8$ (dashed and dotted curves), $\omega=2.6$ (red and green curves), and $\phi_0=0.45$, $\omega=-10$ (blue and dot-dashed curves).}\label{fig6}
	\end{center}
\end{figure}

\begin{figure}
	\begin{center}
		\includegraphics[width=6.6cm]{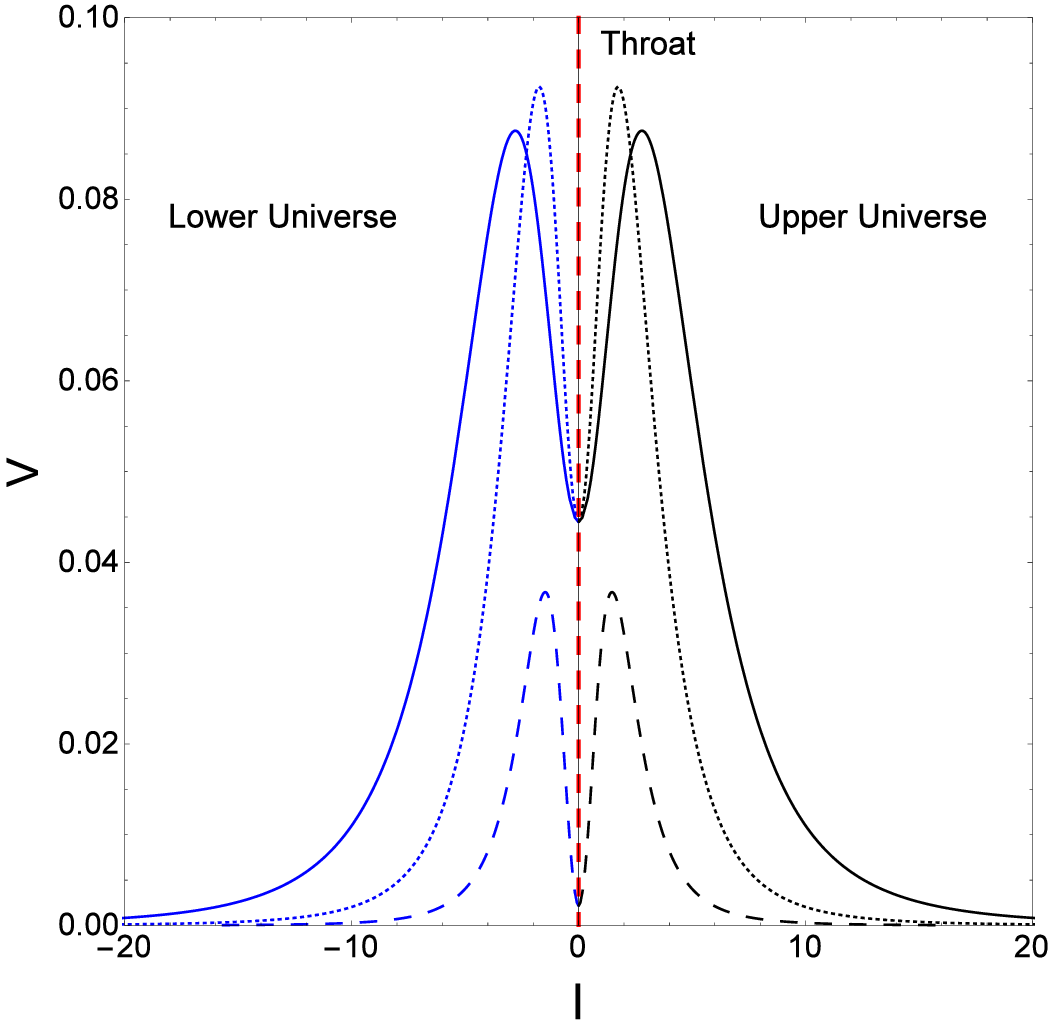}
		\includegraphics[width=6.5cm]{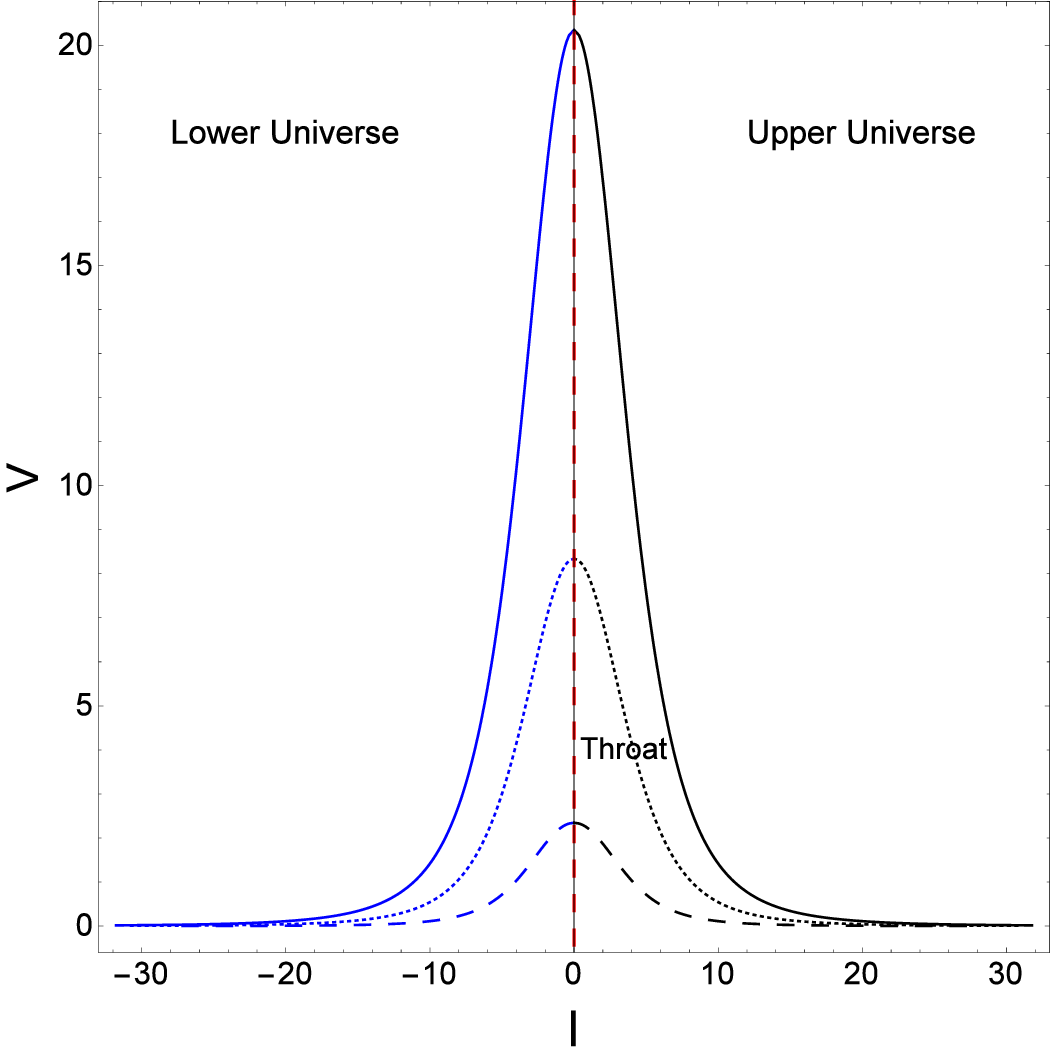}
		\includegraphics[width=6.9cm]{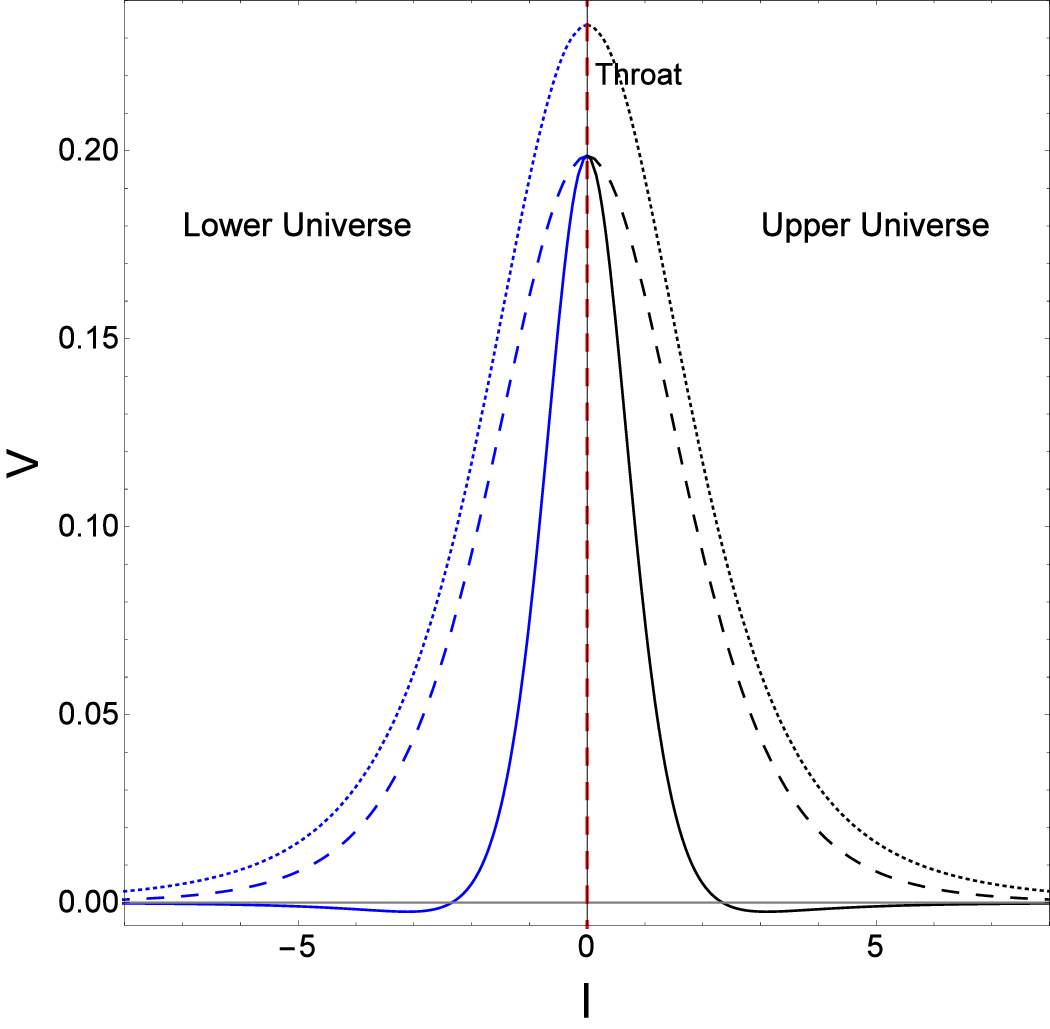}
		\caption{Upper left panel: The behavior of scalar field potential for $\lambda=-\pi^2/1440$, $\phi_0=-0.15$ and $\omega=10$ (solid curves), $\omega=4$ (dotted curves) and $\omega=-1.20$ (dashed curves). Upper right panel: The behavior of scalar field potential for $\lambda=-\pi^2/1440$, $\omega=-10$, $\phi_0=10$ (solid curves), $\phi_0=4$ (dotted curves) and $\phi_0=1$ (dashed curves). Lower panel: The behavior of scalar field potential for $\lambda=(7/8)\times\pi^2/1440$, $\phi_0=0.25$, $\omega=-0.8$ (solid curves) and $\omega=2.6$ (dashed curves) and, $\phi_0=0.45$ and $\omega=-10$ (dotted curves).}\label{fig7}
	\end{center}
\end{figure}

In the present section we proceed to find wormhole solutions assuming a non-vanishing redshift function. We take the functionality of BD scalar field as before and set $m=4$ for scalar Casimir energy between two parallel plates. Then, Eqs.~(\ref{rhoehoc}) and (\ref{bdevol2}) will take the following forms 
\bea
&&4n\phi_0r_0^nr^3(b-r)\left[\Phi^{\prime\prime}+\left(\Phi^{\prime}\right)^2\right]+4n\phi_0r_0^nr^2\Phi^{\prime}\left[\f{r}{2}b^\prime+\left(n\omega+\f{3}{2}\right)b-(n\omega+2)r\right]+2r^{n+5}V^\prime\nn&&2n\phi_0r_0^nr^2\left(1+\f{n}{2}(2\omega+1)\right)b^\prime-n^2\phi_0rr_0^n\Bigg[n\left(3\omega+2\right)\left(b-r\right)+\left(2\omega+1\right)\left(2r-b\right)\Bigg]\nn&&+nr^n\left(Vr^4+16\pi\lambda\right)=0,\label{Phinon}\\
&&n\phi_0r_0^n\left[n(\omega+2)-1\right]\f{b}{r^{n+3}}-\phi_0r_0^n(n-2)\f{b^\prime}{r^{n+2}}+n\phi_0\left[2-n\left(\omega+2\right)\right]\f{r_0^n}{r^{n+2}}-\f{16\pi\lambda}{r^4}-V=0.\label{bnon}\nn
\eea
Now, for $n=2$ Eq.~(\ref{bnon}) reduces to an algebraic equation that can be easily solved for the scalar field potential with the solution given as
\be\label{Vsolnon}
V(r)=2\phi_0r_0^2(2\omega+3)\f{b(r)}{r^5}-4\phi_0(1+\omega)\f{r_0^2}{r^5}-\f{16\pi\lambda}{r^4}.
\ee
Substituting the above solution into Eq.~(\ref{Phinon}) and after rearranging the terms we get
\bea\label{eqphinon}
&-&\!\!\!\!2r^2r_0^2\phi_0(r-b)\left[\Phi^{\prime\prime}+\left(\Phi^\prime\right)^2\right]+rr_0^2\phi_0\left[rb^\prime+(4\omega+3)b-4r(1+\omega)\right]\Phi^\prime+rr_0^2\phi_0(4\omega+5)b^\prime\nn&-&\!\!\!\!3r_0^2\phi_0(4\omega+5)b+8r\phi_0(1+\omega)r_0^2+32\pi\lambda r=0.
\eea
The above differential equation can be solved for the shape function with the general solution given by
\bea\label{solshape}
b(r)={\rm exp}\!\!\left[-\int\!\!f(r)dr\right]\left(\int\!\!{\rm exp}\!\!\left[\int\!\!f(r)dr\right]\!\!g(r)dr+{\rm C}_1\right),
\eea
where
\bea
f(r)&=&\f{2r^2\left[\Phi^{\prime\prime}+\left(\Phi^\prime\right)^2\right]+r(4\omega+3)\Phi^\prime-12\omega-15}{r^2\Phi^\prime+(4\omega+5)r},\label{ffunc}\\
g(r)&=&2\f{r^2r_0^2\phi_0\left[\Phi^{\prime\prime}+\left(\Phi^\prime\right)^2\right]+2rr_0^2\phi_0(1+\omega)\Phi^\prime-4(1+\omega)r_0^2\phi_0-16\pi\lambda}{r_0^2\phi_0\left[r\Phi^\prime+4\omega+5\right]},\label{gfunc}
\eea
and ${\rm C}_1$ is an integration constant. In view of Eqs.~(\ref{ffunc}) and (\ref{gfunc}), one can find the redshift function through specifying the functionality of $f(r)$ or $g(r)$. The simple choice $f(r)=0$ leaves us with the following solution
\bea\label{redsolgen}
\Phi(r)=\ln\left[\left(\f{r_0}{r}\right)^{\omega+\f{1}{4}(1+\beta)}{\rm e}^{\Phi_0}+\f{{\rm C}_2\left(r_0^{\f{\beta}{2}}-r^{\f{\beta}{2}}\right)}{\beta r^{\omega+\f{1}{4}(1+\beta)}}\right],~~~~~\beta=\left(16\omega^2+104\omega+121\right)^{\f{1}{2}},
\eea
where the first integration constant has been set according to the condition $\Phi(r_0)=\Phi_0={\rm constant}$. For ${\rm C}_2=0$ we substitute Eq.~(\ref{redsolgen}) back into Eqs.~(\ref{gfunc}), (\ref{solshape}) and (\ref{Vsolnon}) and finally obtain the following solutions for the redshift and shape functions and scalar field potential as
\bea
\Phi(r)=\Phi_0+\left[\omega+\f{1}{4}(1+\beta)\right]\ln\left(\f{r_0}{r}\right),~~~~~~b(r)=b_1r+b_2,~~~~~~~~V(r)=\f{V_1}{r^4}+\f{V_2}{r^5},\label{solbVphifinal}
\eea
where the integration constant ${\rm C}_1$ has been specified through the condition $b(r_0)=r_0$ and
\bea\label{b1b2V1V2}
b_1&=&\f{1}{\gamma}\left[3(4\omega+9)-\beta\right]-\f{128\pi\lambda}{r_0^2\phi_0\gamma},~~~~b_2=\f{1}{r_0^2\phi_0\gamma}\left[128r_0\pi\lambda-8\phi_0r_0^3\right],\\
V_1&=&\f{1}{\gamma}\bigg(r_0^2\phi_0\left[-2\beta+2(28\omega+43)\right]+16\pi\lambda\left[\beta-\left(44\omega+67\right)\right]\bigg),\\
V_2&=&\f{(2\omega+3)}{\gamma}\left[256\pi\lambda r_0-16\phi_0r_0^3\right],~~~~~~~\gamma=12\omega+19-\beta.
\eea
From the second part of the solution Eq.~(\ref{solbVphifinal}) we find that the flare-out condition at the throat and for $r>r_0$ requires $b_1<1$ and $b_2>0$, respectively. Also the condition on Lorentzian signature is preserved in the following way: at the throat we have $b(r_0)=r_0\Rightarrow1-b_1=b_2/r_0$. Since the right hand side is a decreasing function of radial coordinate then for $r>r_0>0$ we always have $1-b(r)/r>0$. Also the proper distance as a function of radial coordinate can be obtained as
\bea\label{proplnon}
l(r)&=&\f{\alpha}{1-b_1}\left[r_0\sqrt{1-b_1-\f{b_2}{r_0}}-r\sqrt{1-b_1-\f{b_2}{r}}\right]\nn&+&\f{\alpha b_2}{2(1-b_1)^{\f{3}{2}}}\ln\left[\f{-b_2+2r_0\left(1-b_1+\sqrt{1-b_1}\sqrt{1-b_1-\f{b_2}{r_0}}\right)}{-b_2+2r\left(1-b_1+\sqrt{1-b_1}\sqrt{1-b_1-\f{b_2}{r}}\right)}\right].
\eea
Now, on substitution of the solution Eq.~(\ref{solbVphifinal}) into the expressions for energy density and pressure profiles, Eqs.~(\ref{rhoex})-(\ref{ptexp}), we get
\bea\label{rhoprptnon}
\rho(r)\!\!&=&\!\!\f{\lambda}{r^4},~~~~~~~~~~p_r(r)=\f{p_1}{r^4}+\f{p_2}{r^5},~~~~~~~~~~p_t(r)=\f{p_3}{r^4}+\f{p_4}{r^5},\nn\rho(r)+p_r(r)\!\!&=&\!\!\f{\lambda+p_1}{r^4}+\f{p_2}{r^5},~~~~~~~~~\rho(r)+p_t(r)=\f{\lambda+p_3}{r^4}+\f{p_4}{r^5},
\eea
where
\bea\label{p1234}
p_1&=&\f{\lambda}{\gamma}\left[\beta-(76\omega+115)\right]+\f{2\phi_0r_0^2(2\omega+3)}{\pi\gamma},~~~~p_2=\f{32\lambda}{\gamma}(2\omega+3)r_0-\f{2r_0^3(2\omega+3)\phi_0}{\pi\gamma},\nn
p_3&=&\f{\lambda}{\gamma}\left[(8\omega+11)\beta+32\omega^2+132\omega+127\right]-\f{\phi_0r_0^2}{4\pi\gamma}\left[(2\omega+3)\beta+8\omega^2+30\omega+27\right],\nn
p_4&=&\f{(2\omega+3)}{4\pi\gamma}(4\omega+13+\beta)\left(\phi_0r_0^3-16\pi\lambda r_0\right).
\eea
Hence, the energy conditions at the throat read
\bea
\rho\Big|_{r=r_0}\!\!\!\!=\f{\lambda}{r_0^4}\geq0,~~~~~\rho(r)+p_r(r)\Big|_{r=r_0}\!\!\!\!=\f{\lambda+p_1}{r_0^4}+\f{p_2}{r_0^5}=0,~~~~~\rho(r)+p_t(r)\Big|_{r=r_0}\!\!\!\!=\f{\lambda+p_3}{r_0^4}+\f{p_4}{r_0^5}\geq0.\label{energysnon}
\eea
Firstly we note that the positivity of energy density at the throat and throughout the spacetime depends on attractive ($\lambda<0$) or repulsive ($\lambda>0$) nature of the Casimir force. Secondly, at the throat, the matter distribution in radial direction behaves like dark energy with linear EoS $p_r(r_0)=-\rho(r_0)$. Such a behavior can also occur in tangential direction if $p_4=-r_0(\lambda+p_3)$ or in terms of the value of BD scalar field at the throat
\be\label{omegadarkt}
\phi_0=\f{\pi\lambda}{(2\omega+3)r_0^2}\left[107\omega+2\left(5+\sqrt{121+8\omega(13+2\omega)}\right)\right].
\ee
We note that the condition $\beta^2>0$ restricts the values of BD parameter as
\be\label{omegacond}
\omega<-\f{1}{4}\left(13+4\sqrt{3}\right)\lor\omega>\f{1}{4}\left(-13+4\sqrt{3}\right).
\ee
This condition along with the flare-out and energy conditions put limitations on permitted values of the BD parameter and throat radius. The left panel {in} Fig.~(\ref{fig4}) presents the allowed regions of the parameters $\{\omega,r_0\}$ for the case of attractive Casimir effect, according to the conditions stated above. As we observe, wormhole configurations with different throat radii require different allowed values for the BD coupling parameter. The right panel shows the behavior of NEC in radial (family of black curves) and tangential (family of blue curves) directions where it is seen that though the NEC is satisfied at the throat it is violated for $r>r_0$, see the solid and dotted curves. This scenario occurs for $\phi_0<0$. However, for $\phi_0>0$ and $\omega<0$, it is still possible to have NEC satisfied in both directions throughout the spacetime as is seen in Fig.~(\ref{fig5}). The left panel in Fig.~(\ref{fig6}) shows the allowed values of $\{\omega,r_0\}$ parameters for $\lambda>0$ and $\phi_0>0$ and the right panel displays the corresponding behavior of NEC in radial and tangential directions. We observe that this condition can be fulfilled at the throat and for $r>r_0$, see the blue and dot-dashed curves. The case of $\phi_0<0$ does not give physically reasonable values for BD parameter and throat radius. We therefore conclude that depending on model parameters, the NEC and WEC (for $\lambda>0$) can be satisfied at the wormhole throat and throughout the spacetime. The behavior of scalar field potential in terms of the proper length can be obtained using the third part of the solution Eq.~(\ref{solbVphifinal}) along with Eq.~(\ref{proplnon}). Fig.~(\ref{fig7}) shows such behavior for different values of the parameters $\{\lambda,\omega,\phi_0\}$. The upper left panel has been plotted for $\lambda<0$, $\phi_0<0$ and allowed values of BD parameter according to the left panel of Fig.~(\ref{fig4}). We find that in the regions near the throat the BD scalar field encounters a potential barrier that the height of which depends on the value and sign of BD coupling parameter. Hence, for a wormhole configuration with this type of potential, if the scalar field succeeds in crossing the potential peak in the upper universe it may enter the lower one. However, there is a potential well at the throat in which if the scalar field is trapped when the static configuration forms, then the field will stand still between the upper and lower universes. Such a potential well is absent for $\phi_0>0$, see the upper right panel. For this case only a potential barrier exists in front of transmission of the scalar field from upper to lower universes. The height of this potential barrier depends on the values of scalar field at the throat. Also, for $\lambda>0$, (lower panel) the situation is similar to the previous cases except that, $i)$ the WEC is satisfied, see the right panel in Fig.~(\ref{fig6}), $ii)$ in the limit $l\rightarrow0$ the scalar field faces a potential well before encountering the maximum of the potential, see the solid curves.
\par
\begin{figure}
	\begin{center}
		\includegraphics[width=7.7cm]{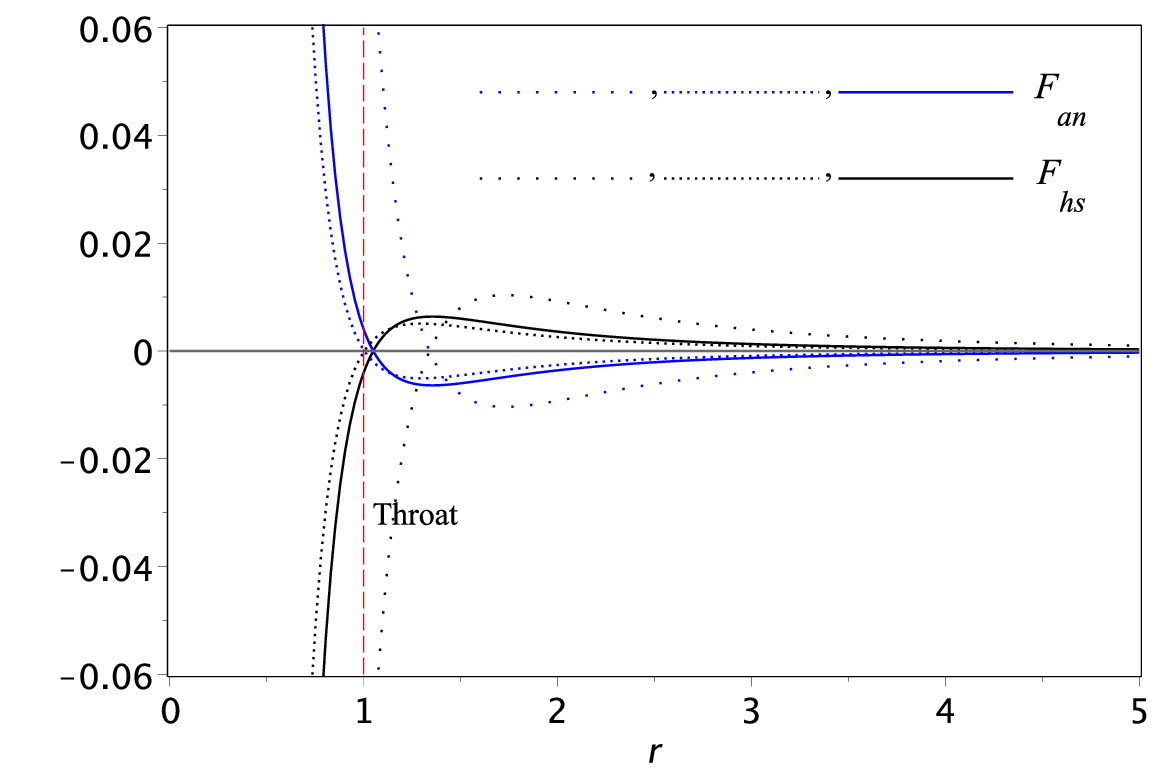}
		\includegraphics[width=7.7cm]{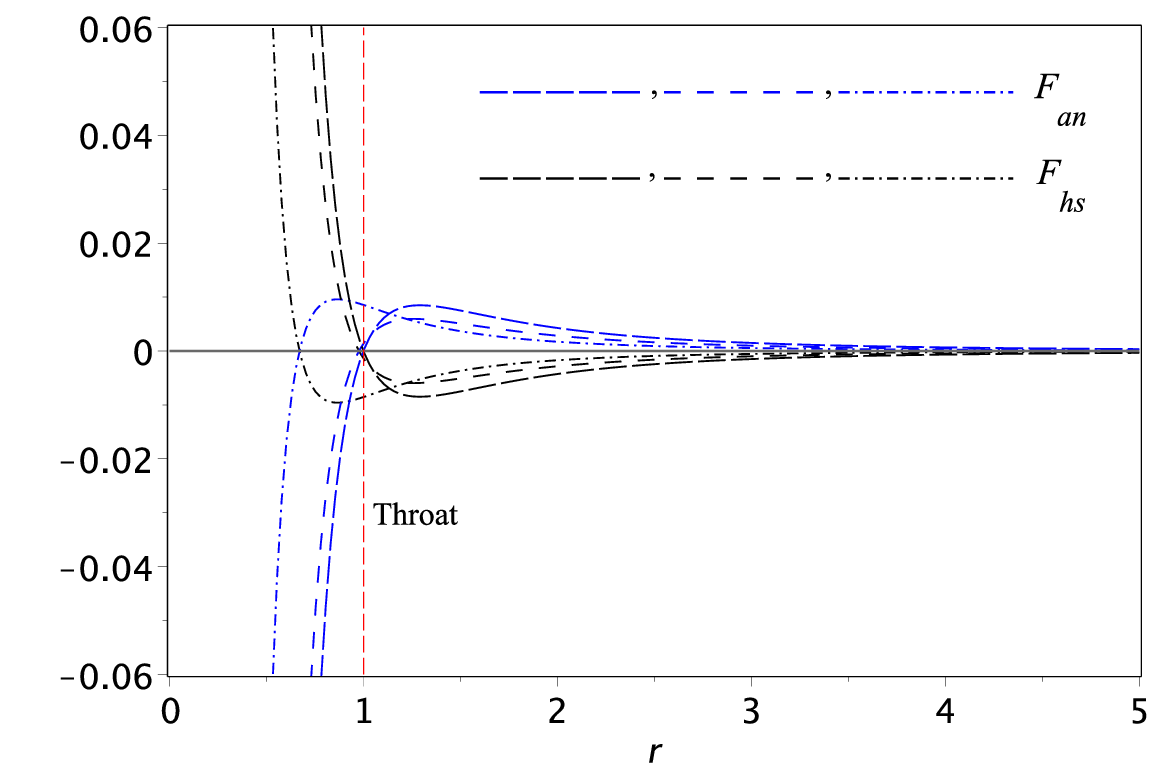}
		\caption{The behavior of anisotropic and hydrostatic forces for $m=4$, $\lambda=-\pi^2/1440$ (left panel) and $\lambda=(7/8)\times\pi^2/1440$ (right panel). The model parameters have been set as of the left panel in Fig.~(\ref{fig1}) with the same line styles.}\label{fig8}
	\end{center}
\end{figure}
\begin{figure}
	\begin{center}
		\includegraphics[width=7.7cm]{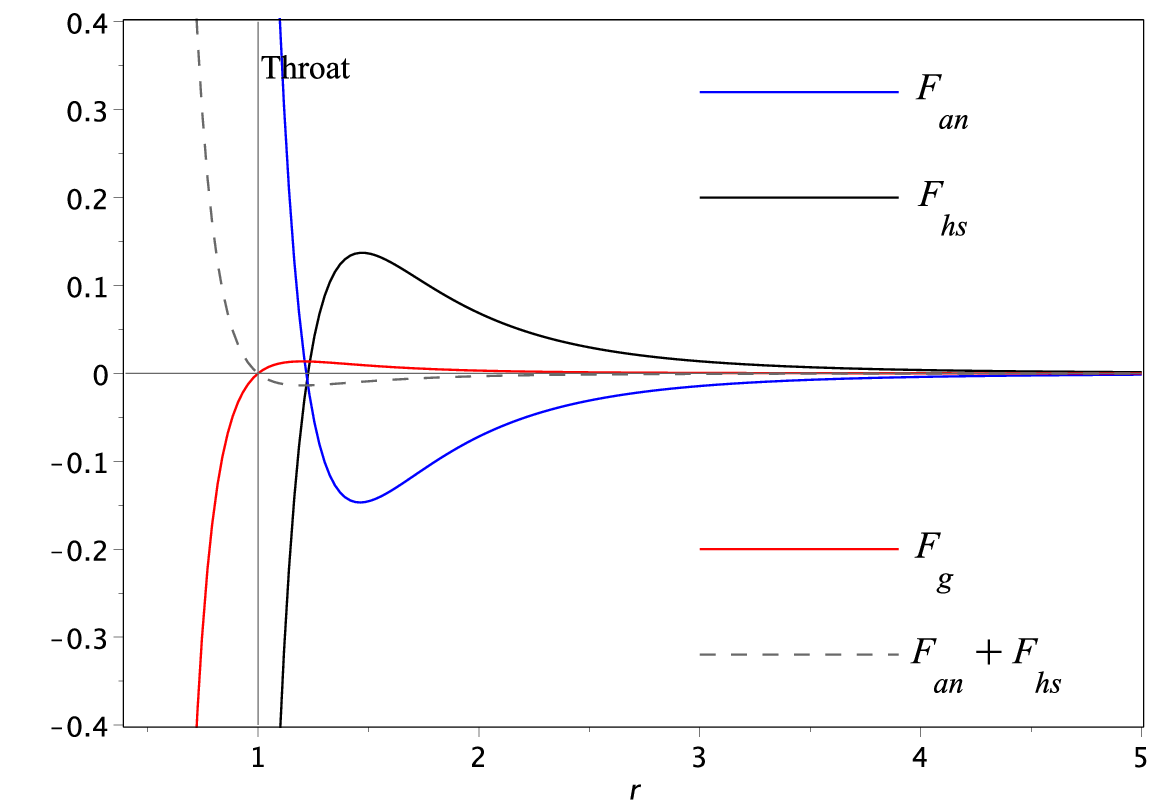}
		\includegraphics[width=7.7cm]{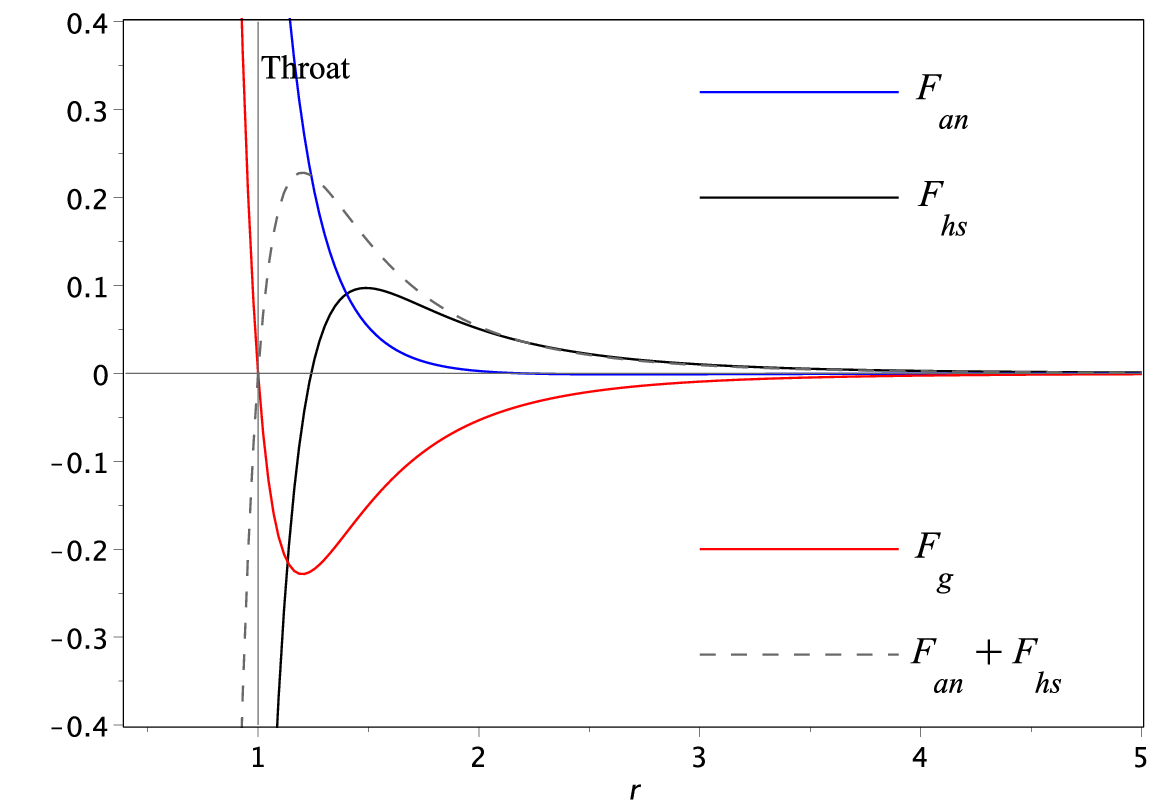}
		\caption{The behavior of anisotropic and hydrostatic forces for $m=4$, $\lambda=-\pi^2/1440$, $\omega=-1.20$ and $\phi_0=-0.15$ (left panel) and, $\omega=-10$ and $\phi_0=10$ (right panel).}\label{fig9}
	\end{center}
\end{figure}
\begin{figure}
	\begin{center}
		\includegraphics[width=7.7cm]{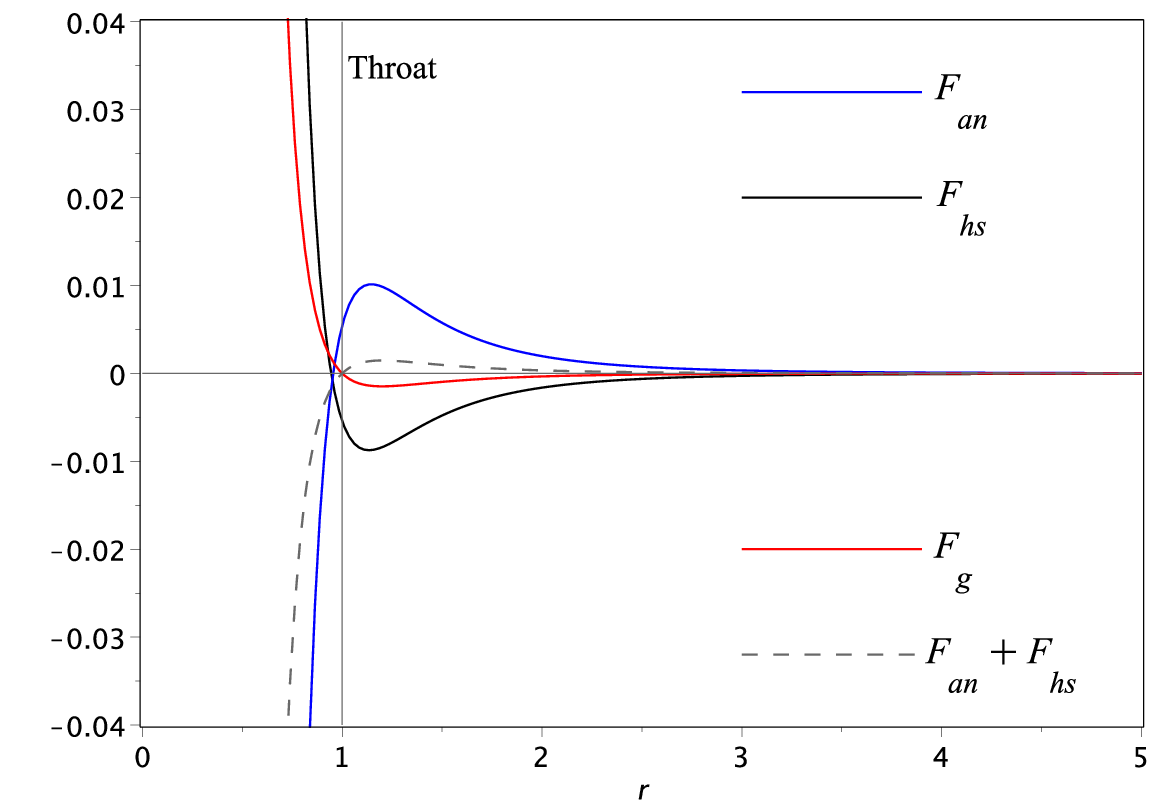}
		\includegraphics[width=7.7cm]{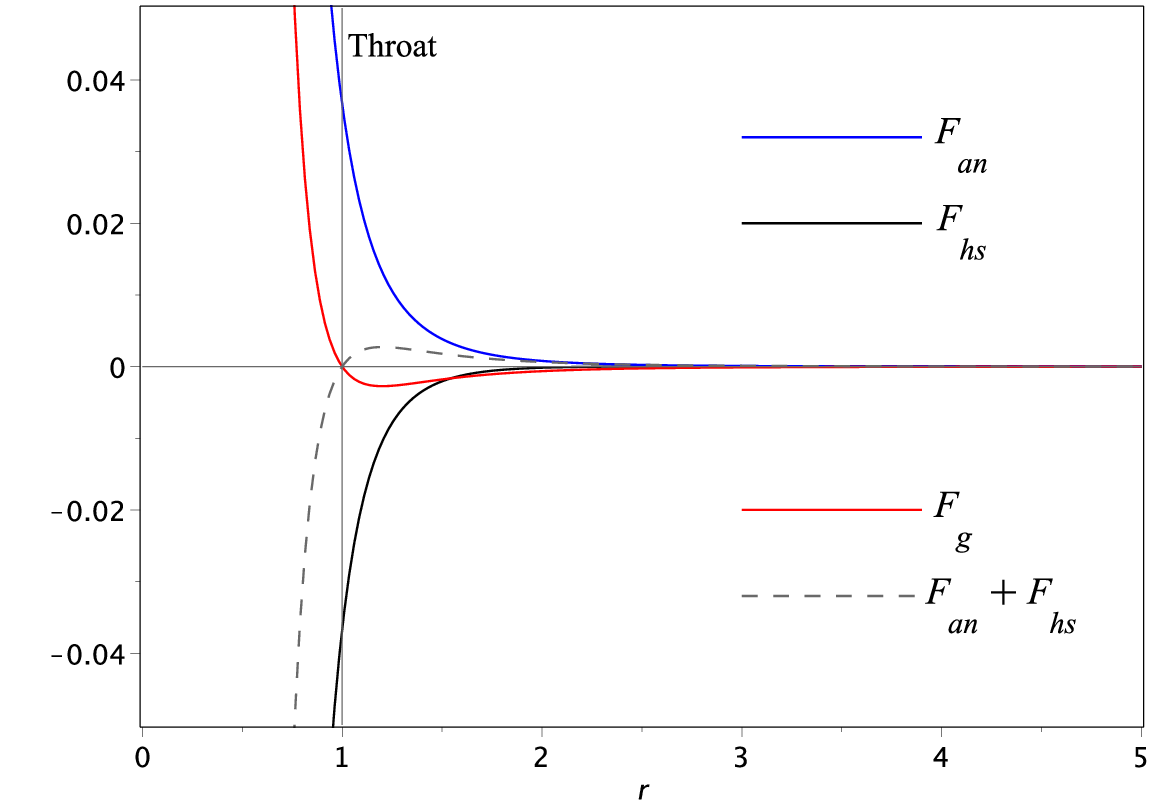}
		\caption{The behavior of anisotropic and hydrostatic forces for $m=4$, $\lambda=(7/8)\times\pi^2/1440$, $\omega=-0.8$ and $\phi_0=0.25$ (left panel) and, $\omega=-10$ and $\phi_0=0.45$ (right panel).}\label{fig10}
	\end{center}
\end{figure}

To close this section let us explore the stability of the obtained solutions using equilibrium conditions. For this purpose we utilize the TOV equation which is given by the conservation equation Eq.~(\ref{conseq}). The equilibrium state of the wormhole structure is determined through the three terms in this equation which are defined as gravitational, hydrostatic and anisotropic forces, respectively
\bea\label{forces}
F_{\rm g}=-\Phi^\prime(\rho+p_r),~~~~~~F_{\rm h}=-\f{dp_r}{dr},~~~~~~F_{\rm an}=\f{2}{r}(p_t-p_r).
\eea
The TOV equation then {reads}
\be\label{forcestov}
F_{\rm g}+F_{\rm h}+F_{\rm an}=0.
\ee
Hence, for the wormhole solutions to be in equilibrium the above equality must hold. From these considerations we can now examine stability of the wormhole solutions discussed above under gravitational, hydrostatic and anisotropic forces. For the case of vanishing redshift we can use the solutions we found in subsection (\ref{redzero}) to get
\bea\label{forcesred0}
F_{\rm an}=\f{(2\omega+3)\left[8\pi\lambda mr_0^{2-m}-\phi_0(m-2)\right]}{2\pi(4\omega+5)(m-2)r^3}+\f{2\phi_0r_0^2(m-2)(2\omega+3)-2\pi\lambda mr^{4-m}(m+8\omega+10)}{2\pi(4\omega+5)(m-2)r^5}.
\eea
The hydrostatic and gravitational forces are found as $F_h=-F_{an}$ and $F_g=0$. In Fig.~(\ref{fig8}) we have sketched the profile of anisotropic (family of blue curves) and hydrostatic (family of black curves) forces where, it is seen that both of these forces show a similar behavior but in opposite direction and hence cancel each other's effect, leaving thus a stable wormhole configuration. Moreover, at the throat we get the anisotropic force as
\bea\label{Fan0}
F_{\rm an}\Big|_{r=r_0}=\f{\phi_0(2\omega+3)r_0^2-8\pi\lambda}{2\pi(4\omega+5)r_0^5},
\eea
from which we find that subject to the inequalities given in Eqs.~(\ref{ineqs4}) and (\ref{ineqs14}), the anisotropic force is always positive at the throat. This implies that the geometry at the throat is repulsive due to the anisotropy of the system, however, these repulsive effects are balanced by the hydrostatic force~\cite{ghsanf}. Finally, for the case of non-constant redshift we have
\bea\label{fanfhsn} 
F_{\rm an}=\f{1}{2\pi\gamma}\left[\f{f_0}{r^5}+\f{f_1}{r^6}\right],~~~~~~~F_{\rm hs}=\f{1}{\pi\gamma}\left[\f{f_2}{r^5}+\f{f_3}{r^6}\right],~~~~~~~F_{\rm g}=\f{1}{2\pi\gamma}\left[\f{f_4}{r^5}+\f{f_5}{r^6}\right],
\eea
where
\bea\label{f123456}
f_0\!\!\!\!&=&\!\!\!\!8\pi\lambda\left[(4\omega+5)\beta+16\omega^2+104\omega+121\right]-\left[\beta(2\omega+3)+8\omega^2+46\omega+51\right]\phi_0r_0^2,\nn
f_1\!\!\!\!&=&\!\!\!\!\left(\beta+4\omega+21\right)(2\omega+3)\left[\phi_0r_0^3-16\pi\lambda r_0\right],~~~~~~
f_2=4\pi\lambda\left(\beta-76\omega-115\right)+8\phi_0(2\omega+3)r_0^2,\nn
f_3\!\!\!\!&=&\!\!\!\!(2\omega+3)\left(160\pi\lambda r_0-10\phi_0r_0^3\right),~~~f_4=(2\omega+3)(\beta+4\omega+1)\left(\phi_0r_0^2-16\pi\lambda\right),~~f_5=-r_0f_4.
\eea
	In Fig.~(\ref{fig9}) we have depicted the behavior of forces given in Eq.~(\ref{fanfhsn}) for $\lambda<0$ and different values of $\{\omega,\phi_0\}$ parameters. Figure (\ref{fig10}) also presents the behavior of these forces for $\lambda>0$. We find that the sum of anisotropic and hydrostatic forces exactly cancel the gravitational force thus leaving a stable wormhole configuration by satisfying the equilibrium condition.
\section{Concluding Remarks}\label{concluding}
Since the work of Morris and Thorne~\cite{mt,mt1}, researches on wormhole structures have opened up a new frontier in theoretical physics. There already exists a large number of research works exploring the possible existence of wormhole geometries in different gravitational theories and under a variety of physical settings. The study of these interesting geometries has made its way into BD theory and much attention has been devoted to this arena since the work of Ellis and Bronnikov~\cite{EllisBronn} and about two decades later the work by Agnese and La Camera~\cite{bd}. Our aim in the present study was to seek for wormhole solutions in BD theory with Casimir energy as the supporting matter. We therefore assumed that this type of matter originates from the vacuum energy density of a scalar field for a configuration of two parallel planes under both Dirichlet and mixed boundary conditions. The former condition corresponds to an attractive Casimir force and the latter to a repulsive one. With the help of these assumptions, two classes of solutions were obtained. In the first one we dealt with zero tidal force solutions, see subsection~(\ref{redzero}), for which we examined the conditions on wormhole geometry as well as energy conditions. Depending on the value of BD scalar field at wormhole throat and the throat radius, we obtained constraints on BD coupling parameter. According to these constraints we found that static spherically symmetric wormhole configurations can exist without violation of NEC and also WEC (for $\lambda>0$). In the second class, as presented in subsection~(\ref{nonzerored}), we investigated wormhole solutions with non-vanishing redshift function and in a similar way to the previous case, we found two-dimensional spaces of model parameters that put constraints on the values of BD coupling parameter as well as the throat radius. These considerations led us to the conclusion that for the solutions with non-constant redshift function, the NEC and WEC (for $\lambda>0$) can be satisfied at the wormhole throat and throughout the spacetime. {We further} studied the behavior of scalar field potential against the proper length. We observed that there may exist potential barriers as well as potential wells at or near the wormhole throat that the height or depth of which depends on model parameters. {Also,} we examined the stability of the solutions through equilibrium condition. This condition expresses the competition between the three forces of hydrostatic, anisotropy and gravity in a static structure with spherical symmetry. We observed that for both classes of solutions, these forces cancel each other at the throat and throughout the spacetime, leaving thus a stable wormhole configuration {for both attractive and repulsive cases}. Finally, as we near to close this report, there remain {a few} points that beg some illustrations: {$i)$ The BD action in Sec.~\ref{BDFES} is given in the Jordan frame in which there is a non-minimal coupling between the scalar field and curvature. As suggested by Dicke~\cite{Dicke1962}, a conformal transformation of the form $\tilde{g}_{\mu\nu}=\phi g_{\mu\nu}$ breaks this non-minimal coupling which can recast the BD theory into the so-called Einstein frame, where the field equations look more tractable. In this frame, the gravitational sector of the theory assumes the familiar form of general relativity and the scalar field contributes through a kinetic term within the action. The matter part of the action also now depends on the scalar field as the EMT undergoes the corresponding conformal transformation. While the conformal transformation technique has been proven as a helpful mathematical tool, its usage has opened a long debate on whether the two conformal frames are physically equivalent or, on the other hand, the question of which frame is physical, see e.g.,~\cite{Faraoni,fujimaeda},\cite{Narayan2016} for recent reviews. In the context of wormhole physics, it has been shown that while wormhole solutions exist in a certain interval of the BD coupling parameter, they do not exist at all in the Einstein frame unless the energy conditions are violated by hand~\cite{bd3}. However, under the light of the Hamiltonian formalism, it is argued that the two frames are not quantum mechanically equivalent~\cite{Gionti2021}. As quantum fluctuations of vacuum manifest themselves through the Casimir effect, then, a question that arises here is that: are Casimir wormholes in Jordan and Einstein frames physically distinct and distinguishable? Our guess that may possibly address this issue is to examine, as it has been done earlier for a massless conformally-coupled scalar field~\cite{Butcher2014}, the Casimir EMT of the BD scalar field for a wormhole spacetime in both Jordan and Einstein frames.} {$ii)$ Boundary properties have been known to play a significant role in Casimir effect since the early days of quantum field theory~\cite{miltonbook,advances}. The imposition of boundary conditions on a quantum field leads to the modification of zero-point energy of the fluctuating field and consequently causes a shift in the vacuum expectation values of physical quantities such as the energy density and stresses. In fact, the confinement of quantum fluctuations leads to the Casimir force which acts on the boundaries that restrict the quantum field~\cite{advances,cassphere1}. The particular features of the resulting vacuum force depend on the nature of the quantum field, the type of spacetime manifold, the boundary geometries and specific boundary conditions imposed on the field~\cite{Sahar2007}. In this sense, the Casimir force can be attractive or repulsive depending on physical properties of the boundaries which are encoded in boundary conditions~\cite{pcp,Miloni1994,Moste1997,Actor1995}. In addition to its fundamental aspects, the Casimir effect has proven its relevance for technical applications to micro- and nano-electromechanical systems in which small components are in close juxtaposition with each other~\cite{MEMS}. It is therefore both fundamentally intriguing and technologically relevant to know that to what extent the electromagnetic Casimir force can be controlled by modifying the physical properties of the environment. For most geometries, the Casimir force between two media separated by vacuum is attractive\footnote{{Although repulsive Casimir force has been reported in e.g., dielectric media~\cite{dielrepforce} and materials with magnetic or chiral response~\cite{repcasforce}, see also~\cite{amoo2011} and references therein.}}, with a magnitude that becomes tangible in the submicron range and rapidly grows in the nanometer range, see e.g.,~\cite{exp2009,exp2017} for experiments on measurement of the Casimir force. In the case of scalar field scenarios, the scalar Casimir effect arises from a fluctuating scalar field, interacting with objects that impose conditions on it; for example imposing Dirichlet or Neumann conditions on the boundary surfaces that confine the scalar field. The former condition corresponds to vanishing of the scalar field at the boundary and the latter specifies the normal derivative of the scalar field at the boundary to be zero or a constant. Also, for the case of a massless scalar field confined between two parallel plates it has been shown that the attractive (repulsive) Casimir force appears for periodic (anti-periodic) boundary conditions~\cite{asorey2013}, see also~\cite{Valu2010} for imposition of the boundary conditions on the confining boundaries.} {$iii)$ The wormhole solutions we obtained in this report have been derived on the assumption that the time-time component of matter EMT obeys the energy density of scalar Casimir effect. However, one may instead utilize the components of the full EMT of the quantized field under consideration in the vacuum state. For example the components of the vacuum EMT of a scalar field are given by, $\langle T_{ij}\rangle=\pi^2/1440a^4{\rm diag}[-1,-3,1,1]$, see e.g.,~\cite{cassphere1,milton2004}. Hence, in an orthonormal reference frame, one may find the shape and redshift functions along with the BD scalar field and its potential through solving the four differential equations (\ref{rhoex})-(\ref{bdevol1}). In the context of BD theory, the solutions that can be obtained in this way represent {\it Casimir wormholes}, as suggested by Garattini~\cite{Garaworm}. These solutions obey the equations of state $p_r=3\rho$ and $p_t=-\rho$ which come out of Casimir EMT $\langle T_{ij}\rangle$. Nevertheless, the scenario we pursued in this work was to explore wormhole solutions for which the energy density mimics that of the associated component of the Casimir EMT. In this regard, one may call them {\it Casimir-like wormholes} in BD theory.} $iv)$ The situation discussed in subsection (\ref{nonzerored}) holds for the special case with $f(r)=0$. One then may be motivated to search for more complicated (and perhaps physically richer) wormhole solutions {assuming} different forms of the functions given in Eqs.~(\ref{ffunc}) and (\ref{gfunc}). $v)$ We restricted ourselves to the case of scalar Casimir energy of two plane-parallel plates. It is however possible to take more complex Casimir {setups} such as, two concentric cylinders~\cite{CASCONCYN}, the case of plate and a cylinder~\cite{cascynp,Emig2006}, {two identical spheres with finite center-to-center separation or a plate and a sphere~\cite{scassphp,CASCONSPH}, two eccentric spheres~\cite{CASCONSPH1}, a massive vector field between two perfectly conducting concentric spherical bodies~\cite{CASCONSPH2},} three-dimensional box~\cite{cassphere1,threeboxcas}, dielectric bodies~\cite{miltonbook,dicas}, {Casimir effect at nonzero temperature~\cite{advances,Castemper}} and other cases~\cite{advances,pcp}. {In particular, the case of scalar field confined between closely spaced two concentric spheres with radii $a$ and $b$ ($b>a$) has been studied in~\cite{ozsan2012} where it is shown that the scalar Casimir energy coincides with the case of parallel plates for a scalar field in the small separation limit, i.e., when $b\rightarrow a$.} In these situations the vacuum energy density of the involved field depends on the radius and length as well as distance between the objects, dielectric permittivity and other parameters. Consequently, investigating the wormhole structures that may arise from such Casimir configurations can provide us with more interesting and exciting results on Lorentzian wormholes in BD theory. However, dealing with these issues is beyond the scope of the present article and future studies will be reported as independent works.
\section{Acknowledgements} 
We would like to appreciate the anonymous referees for providing useful and constructive comments that helped us to improve the original version of our manuscript.


\begin{thebibliography}{99}
\bibitem{BDTH} C. Brans and R. H. Dicke, Phys. Rev. {\bf 124}, 925 (1961).
\bibitem{FLoboBook} F. S. N. Lobo (Editor), \lq{}\lq{}{\it Wormholes, Warp Drives and Energy Conditions,}\rq{}\rq{} Springer (2017).
\bibitem{khu} M. Visser, \lq{}\lq{}{\it Lorentzian Wormholes: From Einstein to Hawking}\rq{}\rq{} (AIP, Woodbury, USA, 1995).
\bibitem{Flamm} L. Flamm, Phys. Z. {\bf 17} 448 (1916);\\G. W. Gibbons, Editorial note to: Ludwig Flamm, Republication of: contributions to Einsteins theory of gravitation. Gen. Relativ. Gravit. {\bf 47}, 72 (2015).
\bibitem{ERose} A. Einstein and N. Rosen, Phys. Rev. {\bf 48}, 73 (1935);\\D. R. Brill and R. W. Lindquist, Phys. Rev. {\bf 131}, 471 (1963).
\bibitem{misner-wheeler} C. W. Misner and J. A. Wheeler, Ann. Phys. {\bf 2}, 525 (1957); \\C. W. Misner, Phys. Rev. {\bf 118}, 1110 (1960).
\bibitem{Wheelerworm} J. A. Wheeler, Ann. Phys. {\bf 2}, 604 (1957);\\ J. A. Wheeler, Geometrodynamics (Academic, New York, 1962).
\bibitem{misnerwheelerworks} J. A. Wheeler, Phys. Rev. {\bf 97}, 511 (1955).
\bibitem{FulWheel} R. W. Fuller and J. A. Wheeler, Phys. Rev. {\bf 128}, 919 (1962).
\bibitem{GerochJMath} R. P. Geroch, J. Math. Phys. {\bf 8}, 782 (1967).
\bibitem{hisworm} F. S. N. Lobo, Int. J. Mod. Phys. D, {\bf 25}, 1630017 (2016).
\bibitem{mt} M. S. Morris and K. S. Thorne, Am. J. Phys. {\bf 56}, 395
(1988).
\bibitem{mt1} M. S. Morris, K. S. Thorne and U. Yurtsever, Phys. Rev. Lett. {\bf 61}, 1446 (1988).
\bibitem{khu1} D. Hochberg and M. Visser, Phys. Rev. D {\bf 56}, 4745 (1997).
\bibitem{MarcoChianese2017} S. W. Hawking, Phys. Rev D {\bf 46}, 603 (1992);\\ 
E. Poisson and M. Visser, Phys. Rev. D {\bf 52}, 7318 (1995);\\
M. Chianese, E. Di Grezia, M. Manfredonia and G. Miele, Eur. Phys. J. Plus {\bf 132}, 164 (2017).
\bibitem{phantworm} F. S. N. Lobo, Phys. Rev. D {\bf 71}, 124022 (2005); \\P. K. F. Kuhfittig, Class. Quant. Grav. {\bf 23}, 5853 (2006); \\F. S. N. Lobo, F. Parsaei and N. Riazi, Phys. Rev. D {\bf 87}, 084030 (2013);\\
Y. Heydarzade, N. Riazi and H. Moradpour, Can. J. Phys. {\bf 93}, 1523 (2015).
\bibitem{intdarksec} V. Folomeev and V. Dzhunushaliev, Phys. Rev. D {\bf 89}, 064002 (2014).
\bibitem{lobocgqreview} F. S. N. Lobo, Classical and Quantum Gravity Research, 1-78, (2008), Nova Sci. Pub. ISBN 978-1-60456-366-5, arXiv:0710.4474 [gr-qc].
\bibitem{Hawevp} L. H. Ford, T. A. Roman, Phys. Rev. D {\bf 53}, 1988 (1996).
\bibitem{Caseffect} H. Epstein, V. Glaser and A. Jaffe, IL Nuovo Cimento, {\bf 36}, 1016 (1965).
\bibitem{negendensqueez} D. Hochberg, T. W. Kephart, Phys. Lett. B {\bf 268}, 377 (1991).
\bibitem{Klinkhammer1991} G. Klinkhammer, Phys. Rev. D {\bf 43}, 2542 (1991).
\bibitem{minexot} M. Visser, Phys. Rev. D {\bf 39}, 3182(R) (1989); Nucl. Phys. B {\bf 328}, 203 (1989);\\S. Kar, N. Dadhich, M. Visser, Pramana {\bf 63}, 859 (2004);\\E. F. Eiroa and C. Simeone, Phys. Rev. D {\bf 71}, 127501 (2005);\\O. B. Zaslavskii Phys. Rev. D {\bf 76}, 044017 (2007);\\M. Bouhmadi-Lopez, F. S. N. Lobo, P. Martin-Moruno, JCAP {\bf 1411},  007 (2014).
\bibitem{modsgrex} F. S. N. Lobo, Class. Quantum Grav. {\bf 25}, 175006 (2008);\\C. G. Boehmer, T. Harko, F. S. N. Lobo, Phys. Rev. D {\bf 85}, 044033 (2012);\\T. Harko, F. S. N. Lobo, M. K. Mak, and S. V. Sushkov, Phys. Rev. D {\bf 87}, 067504 (2013);\\F. Duplessis and D. A. Easson, Phys. Rev. D {\bf 92}, 043516 (2015);\\M. K. Zangeneh, F. S. N. Lobo, M. H. Dehghani, Phys. Rev. D {\bf 92}, 124049 (2015);\\G. U. Varieschi and K. L. Ault, Int. J. Mod. Phys. D {\bf 25} 1650064 (2016);\\M. Hohmann, C. Pfeifer, M. Raidal, H. Veerm\"{a}e, JCAP 1810 (2018) no.10, 003;\\M. Zubair, F. Kousar, S. Bahamonde, Eur. Phys. J. Plus {\bf 133} 523 (2018);\\G. C. Samanta, N. Godani, Eur. Phys. J. C {\bf 79}, 623 (2019);\\A. \"{O}vg\"{u}n, K. Jusufi, I. Sakalli, Phys. Rev. D {\bf 99}, 024042 (2019).
\bibitem{higherdimw} A. Chodos and S. Detweiler, Gen. Rel. Grav. {\bf 14}, 879 (1982);\\G. Cl\'{e}ment, Gen. Rel. Grav. {\bf 16}, 131 (1984);\\A. De Benedictis and A. Das, Nucl. Phys. B {\bf 653}, 279 (2003).
\bibitem{nonsymgr} J. W. Moffat and T. Svoboda, Phys. Rev. D {\bf 44}, 429 (1991).
\bibitem{gmfl} S. H. Mazharimousavi, M. Halilsoy, and Z. Amirabi, Phys.
Rev. D {\bf 81}, 104002 (2010);\\ P. Kanti, B. Kleihaus and J. Kunz, Phys. Rev. D {\bf 85}, 044007 (2012);\\
G. Antoniou, A. Bakopoulos, P. Kanti, B. Kleihaus and Jutta Kunz, arXiv:1904.13091 [hep-th].
\bibitem{braneworm} L. A. Anchordoqui and S. E. Perez Bergliaffa, Phys. Rev. D {\bf 62}, 076502 (2000);\\C. Barcel\'{o} and M. Visser, Nucl. Phys. B {\bf 584}, 415 (2000);\\K. A. Bronnikov and S.-W. Kim, Phys. Rev. D {\bf 67}, 064027 (2003).
\bibitem{LOVEWORM} G. Dotti, J. Oliva, R. Troncoso, Phys. Rev. D {\bf 75}, 024002 (2007);\\
H. Maeda, M. Nozawa, Phys. Rev. D {\bf 78}, 024005 (2008);\\M. H. Dehghani and Z. Dayyani, Phys. Rev. D {\bf 79}, 064010 (2009);\\M. R. Mehdizadeh and F. S. N. Lobo,  Phys. Rev. D {\bf 93}, 124014 (2016).
\bibitem{fr} N. Furey and A. DeBenedictis, Class. Quantum Grav. {\bf 22}, 313 (2005);\\ F. S. N. Lobo and M. A. Oliveira, Phys. Rev. D {\bf 80}, 104012 (2009); \\A. De Benedictis, D. Horvat, Gen. Relat. Gravit. {\bf 44}, 2711 (2012);\\
M. Sharif and I. Nawazish, Annals of Physics, {\bf 389}, 283 (2018);\\O. Sokoliuk, S. Mandal, P. K. Sahoo, A. Baransky, Eur. Phys. J. C {\bf 82}, 280 (2022).
\bibitem{ecworm} K. A. Bronnikov, A. M. Galiakhmetov, Grav. Cosmol, {\bf 21}, 283 (2015); Phys. Rev. D {\bf 94}, 124006 (2016);\\ M. R. Mehdizadeh, A. H. Ziaie, Phys. Rev. D {\bf 95}, 064049 (2017); Phys. Rev. D {\bf 99}, 064033 (2019).
\bibitem{Garcia-Lobo} N. M. Garcia and F. S. N. Lobo, Phys. Rev. D {\bf 82}, 104018 (2010);\\ M. Zubair, S. Waheed and Y. Ahmad, Eur. Phys. J. C {\bf 76}, 444 (2016);\\ R. Solanki, Z. Hassan, P. K. Sahoo, Chin. J. Phys. {\bf 85} 74 (2023).
\bibitem{rastallworm} H. Moradpour, N. Sadeghnezhad and S. H. Hendi, Can. J. Phys. {\bf 95}, 1257 (2017);\\S. Halder, S. Bhattacharya, and S. Chakraborty, Mod. Phys. Lett. A {\bf 34}, 1950095 (2019);\\I. P. Lobo, M. G. Richarte, J. P. Morais Graça, H. Moradpour, Eur. Phys. J. Plus {\bf 135} 550 (2020);\\Y. Heydarzade, M. Ranjbar, Eur. Phys. J. Plus {\bf 138}, 703 (2023).
\bibitem{otherworms} R. Shaikh, Phys. Rev. D {\bf 92}, 024015 (2015);\\ F. Rahaman, N. Paul, A. Banerjee, S. S. De, S. Ray and A. A. Usmani, Eur. Phys. J. C {\bf 76}, 246 (2016);\\ P. H. R. S. Moraes, P. K. Sahoo, Phys. Rev. D {\bf 96}, 044038 (2017);\\ M. G. Richarte, I. G. Salako, J. P. Morais Graca, H. Moradpour, and A. ovgun, Phys. Rev. D {\bf 96}, 084022 (2017);\\K. Jusufi, N. Sarkar, F. Rahaman, A. Banerjee and S. Hansraj, Eur. Phys. J. C {\bf 78} 349 (2018);\\S. Kiroriwal, J. Kumar, S. K. Maurya, S. Chaudhary, Eur. Phys. J. C {\bf 84}, 414 (2024).
\bibitem{Garaworm} R. Garattini, Eur. Phys. J. C {\bf 79}, 951 (2019).
\bibitem{GUPCAS} K. Jusufi, P. Channuie, M. Jamil, Eur. Phys. J. C {\bf 80}, 127 (2020);\\D. Samart, T. Tangphati, P. Channuie, Nuc. Phys. B {\bf 980}, 115848 (2022);\\ Z. Hassan, S. Ghosh, P. K. Sahoo, V. S. H. Rao, Gen. Relativ. Grav. {\bf 55} 90 (2023).
\bibitem{CASWMODG} S. K. Tripathy, Phys. Dark Univ., {\bf 31}, (2021) 100757;\\O. Sokoliuk, A. Baransky, P. K. Sahoo, Nuc. Phys. B, {\bf 930} 115845 (2022);\\Z. Hassan, S. Ghosh, P. K. Sahoo, K. Bamba, Eur. Phys. J. C {\bf 82}, 1116 (2022).
\bibitem{YukCas} R. Garattini, Eur. Phys. J. C {\bf 81}, 824 (2021);\\P. H. F. Oliveira, G. Alencar, I. C. Jardim, R. R. Landim, Symmetry {\bf 15}, 383 (2023).
\bibitem{YMCASW} A. C. L. Santos, C. R. Muniz, R. V. Maluf, JCAP 09 (2023) 022.
\bibitem{Casotherworks} R. Avalos, E. Fuenmayor, E. Contreras, Eur. Phys. J. C, {\bf 82}, 420 (2022);\\ P. H. F. Oliveira, G. Alencar, I. C. Jardim, R. R. Landim, Mod. Phys. Lett. A. {\bf 37}, 2250090 (2022);\\ W. Javed, A. Hamza, Ali \"{O}vg\"{u}n, Mod. Phys. Lett. A. {\bf 35}, 2050322 (2020).
Blagojevi\'{c}, F. W. Hehl, Gauge Theories of Gravitation a Reader with Commentaries, reprint edn. (World Scientific Pub Co Inc/Imperial College Press, London, 2013).
\bibitem{bd} A. G. Agnese and M. La Camera, Phys. Rev. D {\bf 51}, 2011 (1995).
\bibitem{bd1} K. K. Nandi, A. Islam, and J. Evans, Phys. Rev. D {\bf 55}, 2497 (1997).
\bibitem{brans1962} C. H. Brans, Phys. Rev. {\bf 125}, 2194 1962.
\bibitem{bd2} L. A. Anchordoqui, S. P. Bergliaffa, and D. F. Torres, Phys. Rev. D {\bf 55}, 5226 (1997).
\bibitem{bd3} K. K. Nandi, B. Bhattacharjee, S. M. K. Alam, and J. Evans, Phys. Rev. D {\bf 57}, 823 (1998).
\bibitem{bd3rev} A. Bhattacharya, R. Izmailov, E. Laserra, K. K. Nandi, Class. Quant. Grav. {\bf 28}, 155009 (2011).
\bibitem{bd4} F. He and S.-W. Kim, Phys. Rev. D {\bf 65}, 084022 (2002).
\bibitem{bd5} R. Shaikh and S. Kar, Phys. Rev. D {\bf 94}, 024011 (2016). 
\bibitem{bd6} A. Bhattacharya, I. Nigmatzyanov, R. Izmailov, K. K. Nandi, Class. Quant. Grav. {\bf 26} 235017 (2009).
\bibitem{bd7} A. Bhadra, K. Sarkar, D. P. Datta, K. K. Nandi, Mod. Phys. Lett. A {\bf 22}, 367 (2007);\\K. K. Nandi, I. Nigmatzyanov, R. Izmailov, N. G. Migranov, Class. Quant. Grav. {\bf 25} 165020 (2008);\\ F. S. N. Lobo, M. A. Oliveira, Phys. Rev. D {\bf 81}, 067501 (2010);\\P. S. Letelier and A. Wang, Phys. Rev. D {\bf 48}, 631 (1993);\\F. S. Accetta, A. Chodos, Bin Shao, Nuc. Phys. B {\bf 333}, 221 (1990);\\XG. Xiao, B. J. Carr, L. Liu, Gen. Relativ. Gravit {\bf 28}, 1377 (1996);\\L. A. Anchordoqui, A. G. Grunfeld, D. F. Torres, Grav. Cosmol. {\bf 4} 287 (1998).
\bibitem{Faraoni} V. Faraoni, \lq{}\lq{}{\it Cosmology in Scalar Tensor Gravity,}\rq{}\rq{} (Dordrecht: Kluwer Academic, 2004).
\bibitem{fujimaeda} Y. Fujii and K.-ichi Maeda, \lq{}\lq{}{\it The Scalar-Tensor Theory of Gravitation,}\rq{}\rq{} Cambridge University Press (2003).
\bibitem{BDpath} F. Hammad, D. K. Ciftci, V. Faraoni, Eur. Phys. J. Plus {\bf 134}, 480 (2019). 
\bibitem{omega32} M. Salgado, Class. Quant. Grav. {\bf 23}, 4719 (2006).
\bibitem{miltonbook} K. A. Milton, \lq{}\lq{}{\it The Casimir Effect: Physical Manifestations of Zero-point Energy,}\rq{}\rq{} Singapore: World Scientific (2001). 
\bibitem{pcp} M. Bordag, U. Mohideen, V. M. Mostepanenko, Phys. Rept. {\bf 353}, 1 (2001).
\bibitem{advances} M. Bordag, G. L. Klimchitskaya, U. Mohideen, and V. M. Mostepanenko, \lq{}\lq{}{\it Advances in the Casimir Effect,}\rq{}\rq{} Oxford University Press (2009).
\bibitem{spinorcas} K. A. Milton, Ann. Phys., {\bf 150}, 432 (1983);\\E. R. Bezerra de Mello and A. A. Saharian, Class. Quant. Grav. {\bf 23}, 4673 (2006);\\A. Stokes and R. Bennett, Ann. Phys. {\bf 360}, 246 (2015);\\Y. A. Sitenko, J. Phys. Conf. Series {\bf 670}, 012048 (2016);\\G. Fucci and C. R. Sancho, J. Phys. A: Math. Theor. {\bf 56}, 265201 (2023).
\bibitem{cuboidcas} S. G. Mamaev and N. N. Trunov, Theor. Math. Phys. {\bf 38}, 228 (1979).
\bibitem{emwedge} D. Deutsch and P. Candelas, Phys. Rev. D {\bf 20}, 3063 (1979).
\bibitem{sfwedge} J. S. Dowker and G. Kennedy, J. Phys. A: Math. Gen. {\bf 11}, 895 (1978).
\bibitem{cassphere} K. Olaussen and F. Ravndal, Nucl. Phys. B {\bf 192}, 237 (1981).
\bibitem{cassphere1} V. M. Mostepanenko and N. N. Trunov, Sov. Phys. Usp. {\bf 31}, 965 (1988).
\bibitem{scassphp} A. Bulgac, P. Magierski, A. Wirzba, Phys. Rev. D {\bf 73}, 025007 (2006).
\bibitem{cynshellcas} L. L. De Raad, Jr., K. A. Milton, Annals of Physics, {\bf 136}, 229 (1981);\\A. R. Kitson, A. Romeo, Phys. Rev. D {\bf 74}, 085024 (2006). 
\bibitem{cascynp} L. P. Teo, Phys. Rev. D {\bf 84}, 065027 (2011).
\bibitem{milton2004} K. A. Milton, J. Phys. A {\bf 37}, R209 (2004).
\bibitem{PoissonBook} E. Poisson, \lq{}\lq{}{\it A Relativist's Toolkit: The Mathematics of Black-Hole Mechanics,}\rq{}\rq{} Cambridge University Press (2004).
\bibitem{Hochberg1998} D. Hochberg and M. Visser, Phys. Rev. D {\bf 58}, 044021 (1998).
\bibitem{Robinbound} A. Romeo and A. A. Saharian, J. Phys. A: Math. Gen. {\bf 35}, 1297 (2002).
\bibitem{milton2003sf} K. A. Milton, Phys. Rev. D {\bf 68}, 065020 (2003).
\bibitem{actor1996} A. A. Actor, I Bender, Fortsch. Phys., {\bf 44}, 281, (1996).
\bibitem{HorRonb} G. T. Horowitz and S. F. Ross, Phys. Rev. D {\bf 56}, 2180 (1997).
\bibitem{m6cas} E. A. Power and T. Thirunamachandran, J. Mol. Struc. (Theochem) {\bf 591}, 19 (2002);\\C. D. Fosco and G. Hansen, Ann. Phys. {\bf 455}, 169388 (2023);\\ P. W. Milonni and M.-Li Shih, Contemp. Phys. {\bf 33}, 313 (1992).
\bibitem{omm1ls} C. G. Callan, D. Friedan, E. J. Martinez, and M. J. Perry, Nucl. Phys. B {\bf 262}, 593
(1985);\\E. S. Fradkin and A. A. Tseytlin, Nucl. Phys. B {\bf 261}, 1 (1985);\\D. Blaschke and M. P. Dabrowski, Entropy {\bf 14}, 1978 (2012).
\bibitem{ghsanf} T. A. Roman, Phys. Rev. D, {\bf 47}, 1370 (1993);\\ F. S. N. Lobo, Phys. Rev. D, {\bf 75}, 024023 (2007).
\bibitem{EllisBronn} H. G. Ellis, J. Math. Phys. {\bf 14}, 104 (1973); Gen. Relativ. Gravit. {\bf 10}, 105 (1979);\\K. A. Bronnikov, Acta Phys. Polon. B {\bf 4}, 251 (1973).
\bibitem{Dicke1962} R. H. Dicke, Phys. Rev. {\bf 125}, 2163 (1962).
\bibitem{Narayan2016} N. Banerjee and B. Majumder, Phys. Lett. B {\bf 754}, 129 (2016).
\bibitem{Gionti2021} G. Gionti S. J., Phys. Rev. D, {\bf 103}, 024022 (2021);\\ M. Galaverni and G. Gionti S. J., Phys. Rev. D {\bf 105}, 084008 (2022).
\bibitem{Butcher2014} L. M. Butcher, Phys. Rev. D {\bf 90}, 024019 (2014).
\bibitem{Sahar2007} V. V. Nesterenko, G. Lambiase, G. Scarpetta, Riv. Nuovo Cim. {\bf 27}, 1 (2004);\\ A. A. Saharian, Eur. Phys. J. C {\bf 52}, 721 (2007). 
\bibitem{Miloni1994} P. Miloni, \lq{}\lq{}{\it The Quantum Vacuum: An Introduction to Quantum Electrodynamics,}\rq{}\rq{} Academic Press, San Diego (1994).
\bibitem{Moste1997} V. M. Mostepanenko and N. N. Trunov, \lq{}\lq{}{\it The Casimir Effect and Its Applications,}\rq{}\rq{} United Kingdom, Clarendon Press (1997).
\bibitem{Actor1995} A. A. Actor, Fortschr. Phys. {\bf 43}, 141 (1995).
\bibitem{MEMS} F. M. Serry, D. Walliser, G. J. Maclay, J. Microelectromech. Syst. {\bf 4}, 193 (1995);\\ E. Buks and M.L. Roukes, Phys. Rev. B {\bf 63}, 033402 (2001);\\H. B. Chan, V. A. Aksyuk, R. N. Kleiman, D. J. Bishop, F. Capasso, Phys. Rev. Lett. {\bf 87}, 211801 (2001);\\G. Palasantzas and J. Th. M. DeHosson Phys. Rev. B {\bf 72}, 121409(R) (2005).
\bibitem{exp2009} G. L. Klimchitskaya, U. Mohideen, and V. M. Mostepanenko, Rev. Mod. Phys. {\bf 81}, 1827 (2009);\\G. L. Klimchitskaya, V. M. Mostepanenko, Contemp. Phys. {\bf 47}, 131 (2006).
\bibitem{dielrepforce} J. N. Munday, F. Capasso, V. A. Parsegian, Nature {\bf 457}, 170 (2009).
\bibitem{repcasforce} O. Kenneth, I. Klich, A. Mann, M. Revzen, Phys. Rev. Lett. {\bf 89}, 033001 (2002);\\ R. Zhao, J. Zhou, T. Koschny, E. Economou, and C. Soukoulis, Phys. Rev. Lett. {\bf 103}, 103602 (2009);\\ R. Zhao, Th. Koschny, E. N. Economou, C. M. Soukoulis, Phys. Rev. B {\bf 81}, 235126 (2010).
\bibitem{amoo2011} V. M. Mostepanenko, J. Phys.: Conference Series, {\bf 161}, 012003 (2009);\\S. I. Maslovski and M. G. Silveirinha, Phys. Rev. A {\bf 83}, 022508 (2011);\\E. Amooghorban, M. Wubs, N. A. Mortensen, F. Kheirandish, Phys. Rev. A {\bf 84}, 013806 (2011).
\bibitem{exp2017} S. Reynaud and A. Lambrecht, Quantum Optics and Nanophotonics {\bf 101}, 407 (2017).
\bibitem{asorey2013} M. Asorey, J. M. M.-Castaneda, Nuc. Phys. B {\bf 874}, 852 (2013).
\bibitem{Valu2010} X.-zhou Li, H.-bo Cheng, J.-ming Li, X.-hua Zhai, Phys. Rev. D {\bf 56}, 2155 (1997);\\M. Asorey, A. Ibort, G. Marmo, Int. J. Mod. Phys. A {\bf 20}, 1001 (2005);\\ M. A. Valuyan and S. S. Gousheh, Int. J. Mod. Phys. A {\bf 25}, 1165 (2010);\\ J. M. M.-Castaneda, L. S.-Sanz, M. Donaire, M. T.-Fraile, Eur. Phys. J. C {\bf 80}, 793 (2020).
\bibitem{CASCONCYN} F. C. Lombardo, F. D. Mazzitelli and P. I. Villar, J. Phys. A: Math. Theor. {\bf 41}, 164009 (2008);\\L. P. Teo, Phys. Rev. D {\bf 84}, 065027 (2011);\\M. Bordag and V. Nikolaev, J. Phys. A: Math. Theor. {\bf 42}, 415203 (2009).
\bibitem{Emig2006} T. Emig, R. L. Jaffe, M. Kardar, A. Scardicchio, Phys. Rev. Lett. {\bf 96}, 080403 (2006).
\bibitem{CASCONSPH} J. P. Straley and E. B. Kolomeisky, J. Phys.: Condens. Matter {\bf 29}, 143002 (2017).
\bibitem{CASCONSPH1} L. P. Teo, Phys. Rev. D {\bf 85}, 045027 (2012).
\bibitem{CASCONSPH2} L. P. Teo, Phys. Lett. B {\bf 696}, 529 (2011).
\bibitem{threeboxcas} W. Lukosz, Physica {\bf 56}, 109 (1971);\\J. Ambjorn, S. Wolfram, Ann. Phys. (N.Y.) {\bf 147}, 1 (1983).
\bibitem{dicas} A. Romeo and K. A. Milton, Phys. Lett. B {\bf 621}, 309 (2005);\\In\'{e}s Cavero-Pel\'{a}ez, K. A. Milton Annals Phys., {\bf 320}, 108 (2005).
\bibitem{Castemper} J. Ambjorn and S. Wolfram, Annals Phys., {\bf 147}, 1 (1983);\\ G. Plunien, B. Muller, W. Greiner, Physica A: Statistical Mechanics and its Applications, {\bf 145}, 202 (1987);\\A. C. Tort, F. C. Santos, Phys. Lett. B {\bf 482}, 323 (2000);\\ R. Jauregui, C. Villarreal, S. Hacyan,
Annals Phys. {\bf 321}, 2156 (2006);\\S. C. Lim, L. P. Teo, J. Phys. A: Math. Theor. {\bf 40}, 11645 (2007);\\B. Geyer, G. L. Klimchitskaya, V. M. Mostepanenko, Eur. Phys. J. C {\bf 57}, 823 (2008).
\bibitem{ozsan2012} M. Ozcan, Int. J Mod. Phys. A {\bf 27}, 1250082 (2012).
\end{thebibliography}
\end{document}